\documentclass[twocolumn,trackchanges]{aastex631}
\usepackage{graphicx}
\usepackage{float}
\usepackage{enumitem}
\usepackage{CJK}

\shortauthors{Wang et al.}

\defcitealias{Li_2017}{L17}

\graphicspath{{./}{figures/}}
\begin{document}
\begin{CJK*}{UTF8}{gbsn}

\title{On the origin of the split main sequences of the young massive cluster NGC 1856}

\correspondingauthor{Chengyuan Li}
\email{lichengy5@mail.sysu.edu.cn}

\author[0000-0003-3471-9489]{Li Wang (王莉)}
\affiliation{School of Physics and Astronomy, Sun Yat-sen University, Daxue Road, Zhuhai, 519082, China}
\affiliation{CSST Science Center for the Guangdong-Hong Kong-Macau Greater Bay Area, Zhuhai, 519082, China}

\author[0000-0002-3084-5157]{Chengyuan Li (李程远)}
\affiliation{School of Physics and Astronomy, Sun Yat-sen University, Daxue Road, Zhuhai, 519082, China}
\affiliation{CSST Science Center for the Guangdong-Hong Kong-Macau Greater Bay Area, Zhuhai, 519082, China}

\author[0000-0001-8713-0366]{Long Wang (王龙)}
\affiliation{School of Physics and Astronomy, Sun Yat-sen University, Daxue Road, Zhuhai, 519082, China}
\affiliation{CSST Science Center for the Guangdong-Hong Kong-Macau Greater Bay Area, Zhuhai, 519082, China}

\author[0000-0001-9131-6956]{Chenyu He (贺辰昱)}
\affiliation{School of Physics and Astronomy, Sun Yat-sen University, Daxue Road, Zhuhai, 519082, China}
\affiliation{CSST Science Center for the Guangdong-Hong Kong-Macau Greater Bay Area, Zhuhai, 519082, China}

\author[0000-0002-0716-3801]{Chen Wang (王晨)}
\affiliation{Max Planck Institute for Astrophysics, Garching, Germany}

\begin{abstract}
The detection of split main sequences (MSs) associated with young clusters ($\lesssim$600 Myr) has caught lots of attention. A prevailing scenario is that a bimodality of stellar rotation distribution drives the MS bifurcation. Nevertheless, the origin of the stellar rotation dichotomy remains unclear. Hypotheses involving tidally-locked binaries or blue straggler stars (BSSs) are proposed to explain the observed split MSs. This work examines if the long-term dynamical evolution of star clusters can produce the observed split MSs, through high-performance $N$-body simulation. As a prototype example, the young massive cluster NGC 1856 exhibits an apparent MS bifurcation. Our simulation reports that at the age of NGC 1856, tidally-locked binaries are fully mixed with single stars. This is consistent with the observation that there is no significant spatial difference between blue MS and red MS stars. However, we find that only high mass-ratio binaries can evolve to the tidally-locked phase at the age of the NGC 1856. These tidally-locked binaries will populate a much redder sequence than the MS of single stars rather than a blue MS, which is inconsistent with the hypothesis. The number of tidally-locked binaries cannot account for the observation. Our simulation shows that BSSs produced by binary interactions do populate the blue periphery in the color-magnitude diagram, and their spatial distribution shows a similar pattern of single stars. However, the number of BSSs does not fit the observation. 
\end{abstract}


\section{Introduction} \label{sec:intro}

In recent years, extended main-sequence turnoffs (eMSTOs) appear to be a common feature of massive clusters younger than $\sim$ 2 Gyr (e.g., \citealt{Mackey_2008,milone_multiple_2009,li_extended_2019}). Star clusters younger than $\sim$ 600 Myr ubiquitously exhibit split main sequences (MSs), especially when observing these clusters in the ultraviolet (UV) passbands \citep{10.1093/mnras/sty661}. Although the red MS (rMS) component comprises the majority of MS stars (e.g., \citealt{milone_multiple_2016,10.1093/mnras/stx010}), the numbers of stars in blue MS (bMS) and rMS in some clusters are comparable (e.g., \citealt{Li_2017,sun_tidal-locking-induced_2019}).

Early studies interpreted the eMSTO as the result of prolonged star formation histories (SFHs) alone, with an age spread of several hundred million years among cluster member stars \citep{milone_multiple_2009,goudfrooij_population_2009,goudfrooij_population_2011}. This scenario has led to extensive debates, which included the morphologies of the subgiant branches (SGBs) and red clumps (RCs) \citep{li_exclusion_2014,li_not-so-simple_2014,li_tight_2016,bastian_morphology_2015}, a correlation between the inferred age spread and the age of the cluster \citep{niederhofer_apparent_2015}, and the lack of gas in young massive clusters \citep[YMCs,][]{bastian_constraining_2014}. 
These debates point instead toward a stellar evolutionary effect. It is suggested that stellar rotation plays a significant role in shaping the eMSTOs (e.g., \citealt{bastian_effect_2009,cordoni_extended_2018}) and split MSs (e.g., \citealt{dantona_extended_2015,milone_multiple_2016}). 

The main reason why stars with different rotation rates can mimic the morphologies of split MSs or eMSTOs is the combined effect of gravity darkening and rotational mixing \citep[e.g.,][]{yang_effects_2013}. The former not only includes the inclination effect that makes the fast rotating star appear bluer and brighter when seen from pole-on as compared to equator-on but also the reduction in self-gravity due to rotation will reduce the stellar surface temperature, leading stars to appear redder than their slowly rotating counterparts \citep{10.1093/mnras/84.9.665}. The latter could expand the stellar convective core, prolonging the MS lifetime of fast-rotating stars \citep{doi:10.1146/annurev.astro.38.1.143}, and complicating the morphology of the MSTO. 

\citet{bastian_effect_2009} first proposed the fast stellar rotating scenario. Observational evidence supporting this scenario was provided subsequently by \citet{bastian_high_2017,10.1093/mnras/sty661}. Using H$\alpha$ emission images observed with the {\sl Hubble Space Telescope} (HST), both studies confirmed a high fraction of Be stars around the MSTO regions in young massive Large Magellanic Clouds (LMC) clusters, which populate the reddest part of the eMSTO. Be stars are B-type dwarfs with strong Balmer emission lines, which are believed to have rapid rotation rates \citep{rivinius_classical_2013}. Spectroscopic observations strengthen the rotation scenario to explain the eMSTO and split MSs phenomena. \citet{bastian_extended_2018,10.1093/mnras/stz3583,Marino_2018} found that stars on the red side of the eMSTO have high projected stellar rotation rates ($v$sin$i$) while stars on the blue side have low $v$sin$i$. \citet{marino_different_2018,sun_tidal-locking-induced_2019} corroborated that bMS stars are composed of slow rotators and  rMS stars are mainly rapid rotators. While this scenario successfully explains the eMSTOs and split MSs in most clusters, there are a few exceptions. \citet{milone_multiple_2017} found neither stellar populations with different ages only, nor coeval stellar models with different rotation rates, properly reproducing the observed split MS and eMSTO of NGC 1866. \citet{costa_multiple_2019} suggested that a combination of stellar rotation and age spread is required to fit the observed color-magnitude diagrams (CMD) of NGC 1866. Similar conclusions can be drawn for clusters NGC 1987 and NGC 2249 \citep{goudfrooij_extended_2017}.

While a dichotomy of stellar rotation can explain the bifurcated MS of clusters \citep[e.g.,][]{dantona_extended_2015,sun_tidal-locking-induced_2019}, the underlying parameters which control the dichotomy remains speculative. \citet{10.1093/mnras/staa1332} proposed that such a bimodal rotational distribution depends on the lifetime of proto-stellar discs in pre-main-sequence (PMS) stars. According to their scenario, present-day fast rotators are the progeny of PMS stars with short-lived proto-stellar discs, whereas long surviving time for proto-stellar discs would result in slow rotators. This scenario is yet to be examined, which requires direct measurements of rotations of PMS stars in young star-forming clusters. 

The peculiar blue straggler stars (BSSs), commonly found in dense stellar systems, are located along the blue extension of the MSTO regions in the CMDs of star clusters. One of the main formation channels is binary mass-transfer up to the coalescence of two binary components \citep{hills_stellar_1976,andronov_mergers_2006}. Based on their locations in the CMD, the population of BSSs provides a further clue for the merger origin of the bMS. \citet{wang2022stellar} suggested that bMS stars are slow rotators formed from stellar mergers and rMS are fast rotators formed from disk accretion. They claim that this hypothesis is supported by the unusual flat mass functions of the bMS stars in four clusters. Their scenario assumes that a stellar merger would produce a massive star with a strong magnetic field, leading to a slowly rotating early-type star. Studying the magnetic fields of bMS stars would thus provide deep insight into this hypothesis. 

\citet{d2017stars} suggested that all stars are born as fast rotators and the rotation of some stars is braked due to tidal locking in close binary systems on timescales of a few $\times$ $10^7$ yr, leading to the formation of bMSs. They linked binary interactions to the rotation-spread scenario. 
\citet{sun_tidal-locking-induced_2019} also suggested that the bMS stars in NGC 2287 are likely tidally-locked stars with synchronization timescales shorter than the cluster age. However, special conditions (e.g., as regards the mass-ratio distribution) might be required for this scenario to hold. 
If bMS are populated by tidally-interactive binaries, they should be at least partially hard binaries. 
However, \citet{yang_spatial_2021} suggested that bMS stars in four clusters are partly soft binaries, which is at odds with the rotation spread scenario that interprets them as hard binaries.  
\citet{kamann_exploring_2021} detected similar fractions of binary stars in the slow and rapidly rotating populations in NGC 1850 and concluded that binarity is not a dominant mechanism in forming the observed bimodal rotational distributions.
\citet{he_role_2022} also argued that the high fraction of the tidally interacting binaries is unusual if the stars with low projected rotation rates are tidally-locked binaries. In particular, there is a minor fraction of bMS stars that are too blue to be explained by this scenario alone { \citep{wang2022stellar}}.

Using numerical $N$-body simulations, this study sets out to verify whether the long-term dynamical evolution of star clusters can reproduce the observed photometric split MSs in the CMDs and the spatial distributions of their stellar components. 
The young massive cluster NGC 1856 ($\sim$ 300 Myr, $\sim 10^5 M_{\odot}$ from \citealt{McLaughlin_2005}) in the LMC is an ideal test case for probing the connection between the secular evolution of stellar clusters and the bMS. \citet{10.1093/mnras/stv829} provided the first evidence for the apparent MS bifurcation in NGC 1856. 
\citet[][hereafter \citetalias{Li_2017}]{Li_2017}, found a number ratio of the bMS and rMS stars in NGC 1856 is 34\%/66\%, and speculated that the different stellar rotation scenario would be a convincing explanation of the observed split MSs in NGC 1856. Here, we aim to examine the properties of the tidally-locked binaries and BSSs in an NGC 1856-like cluster and compare them with the observation, using $N$-body numerical simulations incorporating stellar evolutionary models. 

This article is organized as follows. In Section \ref{sec:model}, we describe the $N$-body code {\tt\string PETAR}, and the initial conditions for modeling the evolution of an NGC 1856-like star cluster. In Section \ref{sec:results}, we present the main results of our simulations in detail, followed by a discussion in Section \ref{sec:discussions} and a summary in Section \ref{sec:conclusions}.

 
\section{Methods} \label{sec:model}

\subsection{The $N$-body algorithm, {\tt\string PETAR}}

NGC 1856 is a YMC with a total mass of several $10^5M_{\odot}$ \citep{mclaughlin_resolved_2005}. Directly simulating massive collisional stellar systems, such as globular clusters (GCs), where stars and binaries can have frequent close interactions, is time-consuming due to the large number of stars in clusters. The high-performance $N$-body code, {\tt\string PETAR}\footnote{\url{https://github.com/lwang-astro/PeTar}}, can efficiently simulate the dynamical evolution of massive star clusters with a binary fraction up to unity \citep{10.1093/mnras/staa1915,wang_impact_2022}. {\tt\string PETAR} has been developed based on the framework for developing parallel particle simulation codes ({\tt\string FDPS}), which can avoid expensive computing and achieve high performance on multiple-core computers by using multi-process parallel computing \citep{10.1093/pasj/psw053,10.1093/pasj/psz133,10.1093/pasj/psy062}. 
In this work, we use the {\tt\string PETAR} code to mimic an NGC 1856-like cluster with details of stellar dynamics and evolution. 
\begin{deluxetable}{cc}
 \tablecaption{The initial parameters used for the {\tt\string PETAR} simulations \label{tab:para}}
 \tablewidth{0pt}
 \tablehead{\colhead{Parameters} & \colhead{Initial conditions}}
 \startdata
 cluster total mass & $2\times10^5$ $M_{\odot}$\\
 half-mass radius & 1.4 pc \\
 metallicity & 0.01 \\
 primordial binary fraction & 100\% \\
 mass segregation & -\\
 tidal field & -\\
 \enddata
 \tablecomments{The initial total mass and the initial half-mass radius were estimated empirically from the present-day observed data of \citealt{McLaughlin_2005}.}
 \end{deluxetable}

 To accurately follow the dynamical and stellar evolution of both single stars and binary systems, the recently-updated single and binary stellar evolution codes, {\tt\string SSE} and {\tt\string BSE} \citep{10.1093/mnras/291.4.732,10.1046/j.1365-8711.2000.03426.x,10.1046/j.1365-8711.2002.05038.x,refId1}, are incorporated in {\tt\string PETAR} to simulate the wind mass loss, the type changes of stars, the mass transfer and the mergers of binaries. The update from \citet{refId1} uses semi-empirical stellar wind prescriptions from \citet{Belczynski_2010}, while {\tt\string PETAR} adopt the ``rapid'' supernova model and material fallback from \citet{Fryer_2012}, along with the pulsation pair-instability supernova model \citep{Belczynski2016} for the formation of compact objects. Although the {\tt\string BSE} model for binary evolution includes stellar spins subject to tidal circularization and synchronization, the {\tt\string BSE} is not rigorous in treating merger spins according to our preliminary tests. In this work, we did not analyze stellar spins in the {\tt\string BSE} in the subsequent discussions.

\subsection{Initial Conditions}

The adopted simulation model starts at the gas-free phase, containing only particle-like stellar components. Therefore, the long-term evolution of the star cluster is controlled by the combination of stellar evolution, stellar interaction, and the external Galactic potential. Although our simulation does not consider primordial mass segregation and tidal field, we will discuss the effects of these conditions in Section \ref{sec:discussions}. 

\begin{figure*}[ht!]
  \epsscale{1.1}
 \plottwo{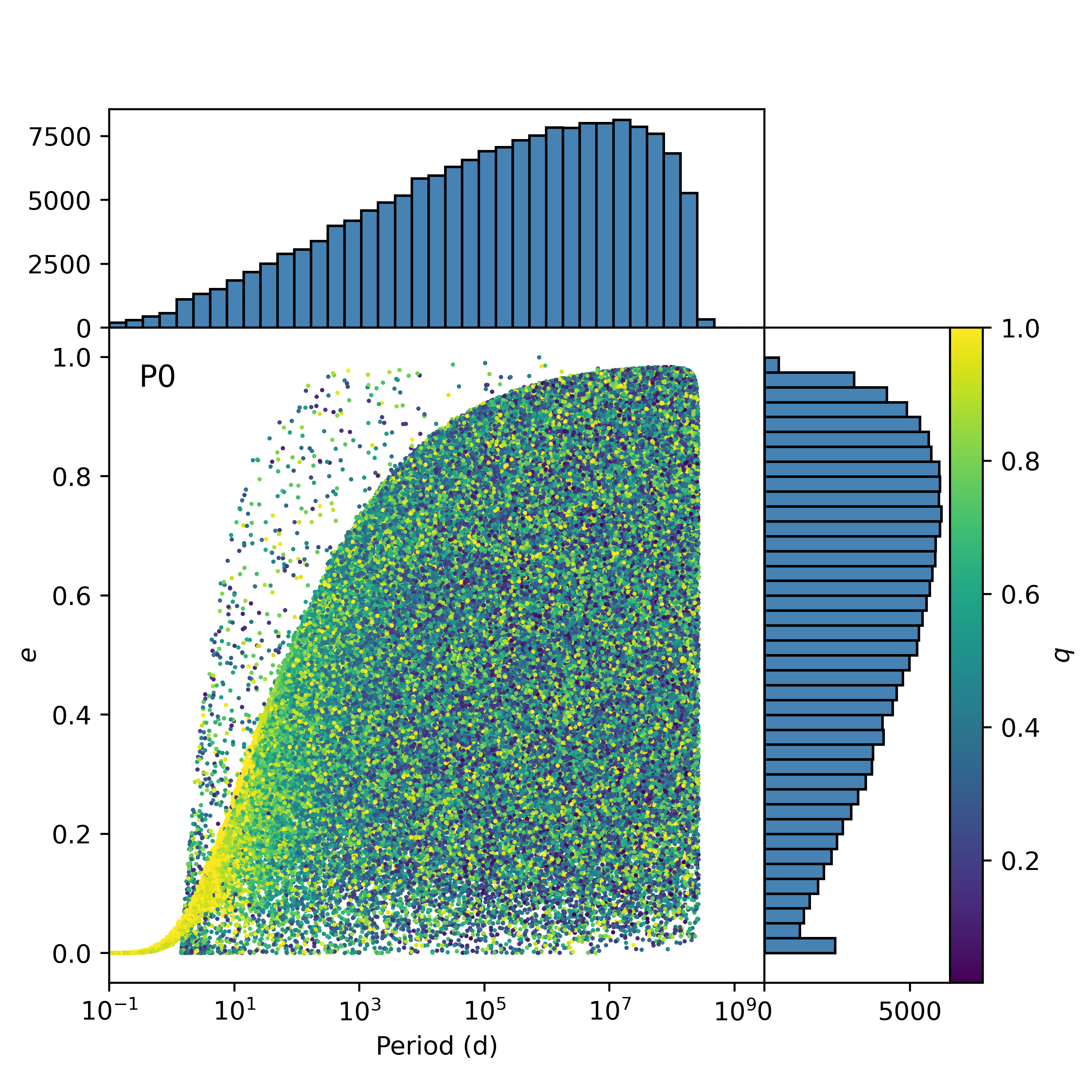}{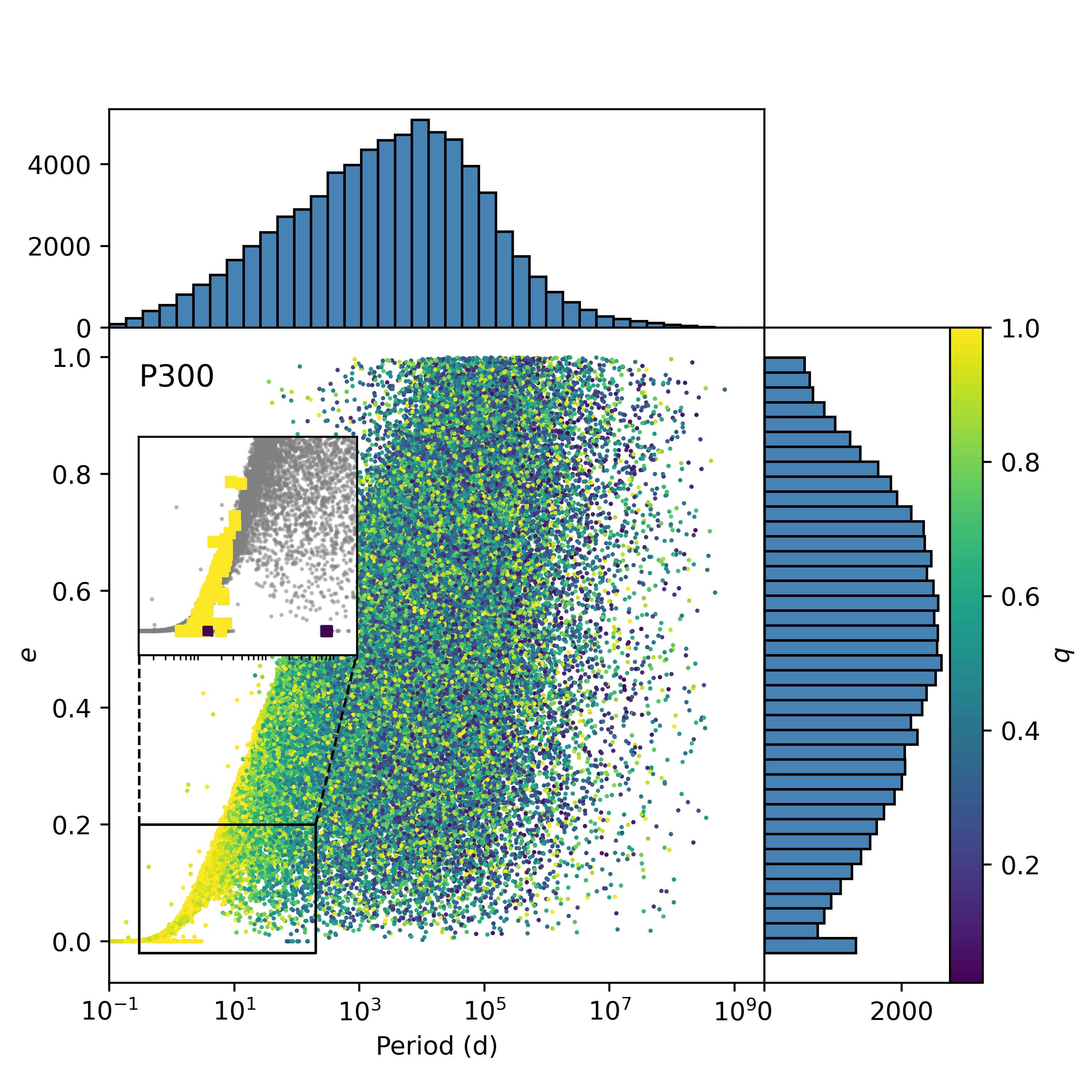}
 \caption{The period$-$eccentricity ($e$) distribution of the P0 (Left) and the P300 (Right). In each panel, dots in the middle plots represent the binaries, the color represents the mass ratio $q$, and the horizontal axes are on logarithmic scales. The upper and the right plots show the histograms of the period and $e$, respectively. The squares in the inset highlight tidally-locked binaries color-coded according to their mass ratios. \label{fig:ae}}
 \end{figure*}

Gas-free star clusters are well characterized observationally by Plummer-like profiles \citep{10.1093/mnras/71.5.460,R2011,R2019}. Our initial setups are empirically motivated and are chosen to reproduce the observed number density profile of NGC 1856. We use the newly updated version of the star cluster initial model generator code {\tt\string MCLUSTER}\footnote{\url{https://github.com/lwang-astro/mcluster}} \citep{10.1111/j.1365-2966.2011.19412.x,10.1093/mnras/sty2232} to generate an NGC 1856-like cluster with initial parameters listed in Table \ref{tab:para}. The initial total mass is $2\times10^5$ $M_{\odot}$, and the initial half-mass radius is 1.4 pc including all stellar components in three-dimensional space, estimated empirically from the present-day observed data. The cluster metallicity is $Z$=0.01 ($Z_{\odot}$=0.02). The initial particle masses were randomly sampled from a Kroupa-like initial mass function (IMF) \citet{10.1046/j.1365-8711.2001.04022.x} with the mass range of 0.08$-$150 $M_{\odot}$. The 3D positions and velocities of stars were randomly sampled from the Plummer density profile \citep{1974A&A....37..183A}.

The simulated cluster evolved to a maximum age of 400 Myr, yielding output snapshots in time intervals of 4 Myr. Hereafter, the names of snapshots combine the prefix "P" with the evolutionary timescales in Myr unit, such as P300. Each snapshot includes age, mass, luminosity, temperature, and { other 31 parameters, which depict the kinematic, evolutionary, and dynamical properties of each star}. The snapshot data are processed with the built-in tools within {\tt\string PETAR} to automatically detect singles and binaries and calculate parameters for binaries, such as eccentricity, period, kinetic energy, and binding energy. In our simulation, the best-fitting model which mimics the real observation of the NGC 1856 is P300. We will introduce our model in the next subsection.

Previous studies suggested that a primordial binary fraction up to 100\% better restores the observed binary fractions inside and outside half-mass radii of GCs \citep{Leigh2015TheSO} and the observed CMDs of GCs \citep{belloni_initial_2017}. The adopted simulation model in this work is thus initialized with a 100\% primordial binary fraction. We generate primordial binaries by randomly sampling orbital parameters from a set of distributions for mass ratio $q$, semi-major axis $a$, and eccentricity $e$, as described below. We define the mass ratio of a binary system consisting of two stars of masses $M_{\rm p}$ and $M_{\rm s}$ as $q = M_{\rm s}/M_{\rm p}$, where $M_{\rm s} \leqslant M_{\rm p}$.

\begin{figure*}[ht!]
  \plottwo{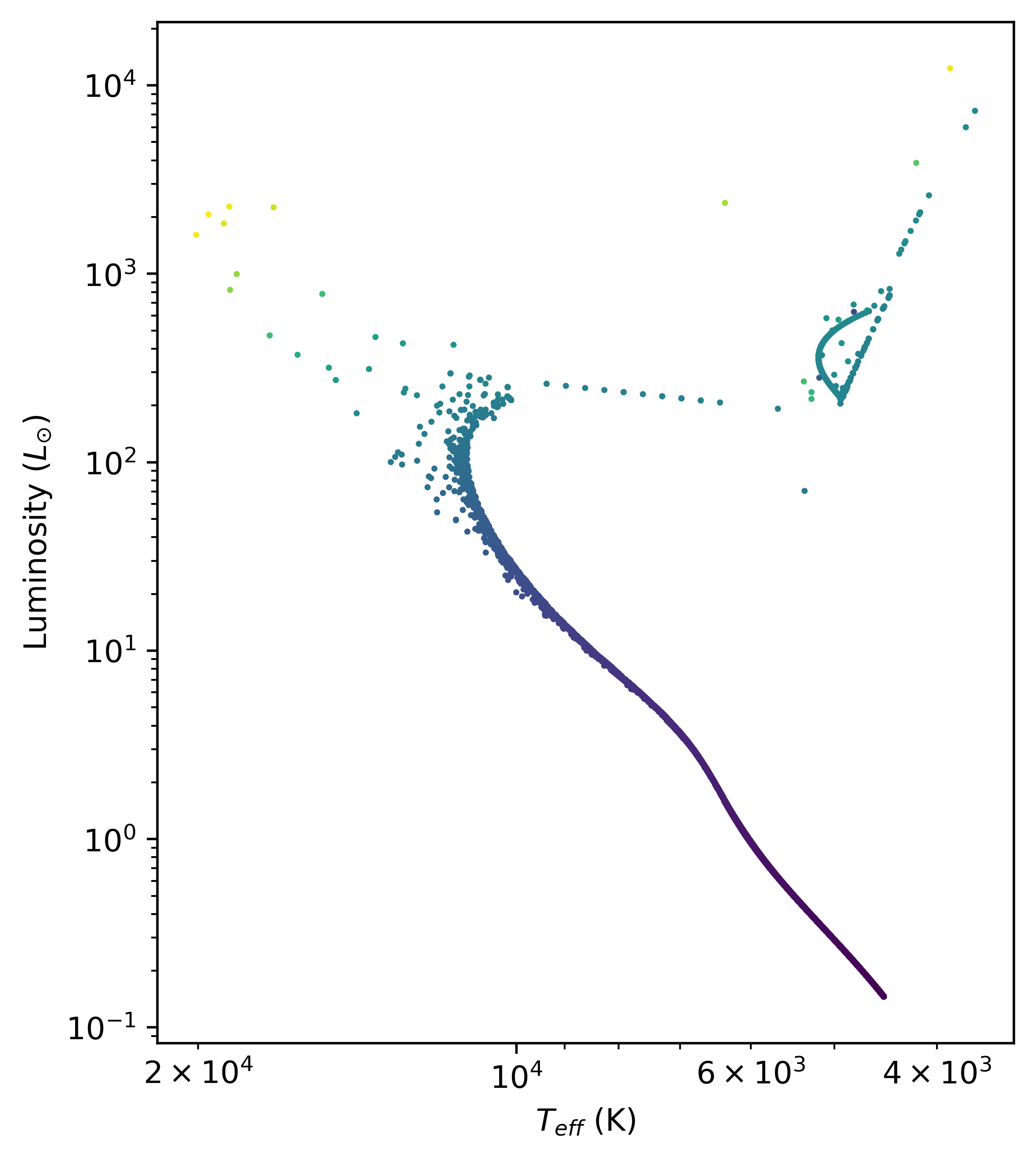}{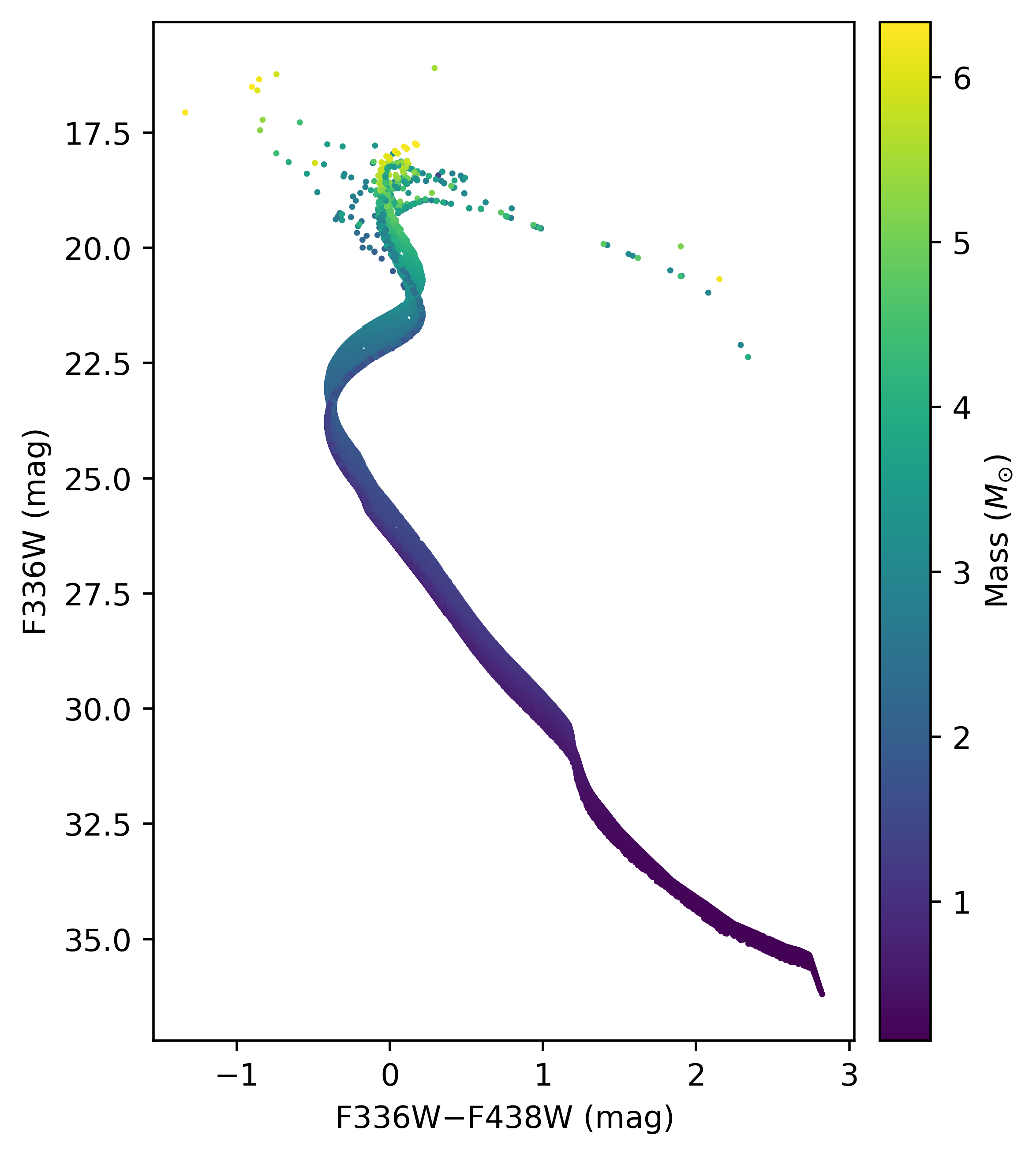}
  \caption{(Left) HRD and (Right) CMD of the P300 model. \label{fig:hrcmd}}
  \end{figure*}

The initial orbital parameter distributions of binary systems impose essential constraints on forming tidally-locked binaries, as shown in Figure \ref{fig:ae}. 
Our simulations follow the initial properties of binaries described in \citet{belloni_initial_2017}. Their numerical simulations with these assumptions provide qualitatively similar good descriptions of observations of Galactic-field late-type binaries, GCs, and open clusters. 
For binaries wtih $M_{\rm p} < 5M_{\odot}$, we apply the initial semi-major axis distribution from \citet{10.1093/mnras/277.4.1507,10.1093/mnras/277.4.1491} and the thermal eccentricity distribution ($f(e)\propto e$, \citealt{10.1093/mnras/173.3.729}). The masses of both binary components are generated through random pairing from the Kroupa IMF. Under the hypothesis that stars form in aggregates of binary systems and that the dynamical evolution of these aggregates leads to the observed properties of binary stars in the Galactic field, the model with these initial setups has a good agreement with the observational data of PMS stars \citep{10.1093/mnras/277.4.1507,10.1093/mnras/277.4.1491}. Period distributions and eccentricity distributions in \citet{sana};\citet{Oh_2015} have been used in our simulation to investigate the properties of primordial binaries with $M_{\rm p} > 5M_{\odot}$. The distributions of orbital parameters of massive binaries are relatively loosely constrained compared to solar-type stars. Moreover, the mass of the secondary star ($M_{\rm s}$) is randomly sampled from a uniform distribution ($0.1 \textless q \textless 1$, \citealt{sana}). These compound setups for massive primordial binaries are based on all relevant binary characteristics measurements in a sample of Galactic massive O-type stars by \citet{sana}.
 
 We compare the distributions of period and $e$ of the P0 model with those of the P300 model in Figure \ref{fig:ae}. For the P0 model in the left panel of Figure \ref{fig:ae}, most primordial binaries have period $ > 10^7$day with the $e$ peaked at 0.8. Compared with P0, the orbital distribution of binaries in P300 tends to be tighter and more circular, and most wide binaries have been dynamically disrupted. That is consistent with the Heggie-Hills law \citep{10.1093/mnras/173.3.729, 1975AJ.....80..809H}: soft binaries tend to be disrupted after numerous encounters with neighboring stars while hard binaries tend to be tighter.

 We convert the luminosity and temperature into the HST's Ultraviolet and Visual channels of the Wide Field Camera 3 (UVIS/WFC3) filters absolute magnitudes by using the PARSEC Bolometric Correction\footnote{\url{https://sec.center/YBC}} ({\tt\string YBC}, \citealt{YBC}). 
 By comparing the observed CMD with the simulated CMD, the distance modulus applied to our simulation is estimated by visual inspection. We display the Hertzsprung Russell diagram (HRD) and observable CMD of the P300 model in Figure \ref{fig:hrcmd}. Given the large distance of the LMC and the crowded environment of the massive cluster NGC 1856, we assume that binaries in our simulation are all unresolved. Because of this, binary systems will appear as single point-like sources in observation. The magnitude of a binary system is
 \begin{equation} 
 m_{\rm b} = -2.5\log(10^{-0.4m_{\rm p}}+10^{-0.4m_{\rm s}})
 \label{eq:binary}
 \end{equation}
 where $m_{\rm p}$ and $m_{\rm s}$ are magnitudes of the primary and secondary stars. The equal-mass binaries will populate a brighter sequence parallel to the MS, with a magnitude difference of $\sim -0.752$ mag. Note that the {\tt\string YBC} does not consider the effects of rotations, the broadening of the MS in the right panel of Figure \ref{fig:hrcmd} is entirely driven by unresolved binaries.

 \subsection{The NGC 1856-like cluster model}

To search for a model that best reproduces the observations of NGC 1856 from \citetalias[]{Li_2017} photometric catalogs, we compare the normalized 2D projected radial number density profile of NGC 1856 with the {\tt\string PETAR} models by only using visible stars more massive than 1.7 $M_{\odot}$, where the lowest mass is from an observational cut-off corresponding to F336W = 20.75 mag. We first define the location where the stellar number density reaches its maximum in the plane of equatorial coordinates ($\alpha_{J2000}$, $\delta_{J2000}$) as the cluster center. 
All subpopulation stars were equally divided into 15 annular rings within 100 arcsec, and each ring has equal width. This selection criterion also applies to the {\tt\string PETAR} model. We calculate the number density of stars in each ring, and its profile can be described by the Elson, Fall, and Freeman (EFF) model \citep{1987ApJ...323...54E},
 
 \begin{figure}[ht!]
 \epsscale{1.18}
 \plotone{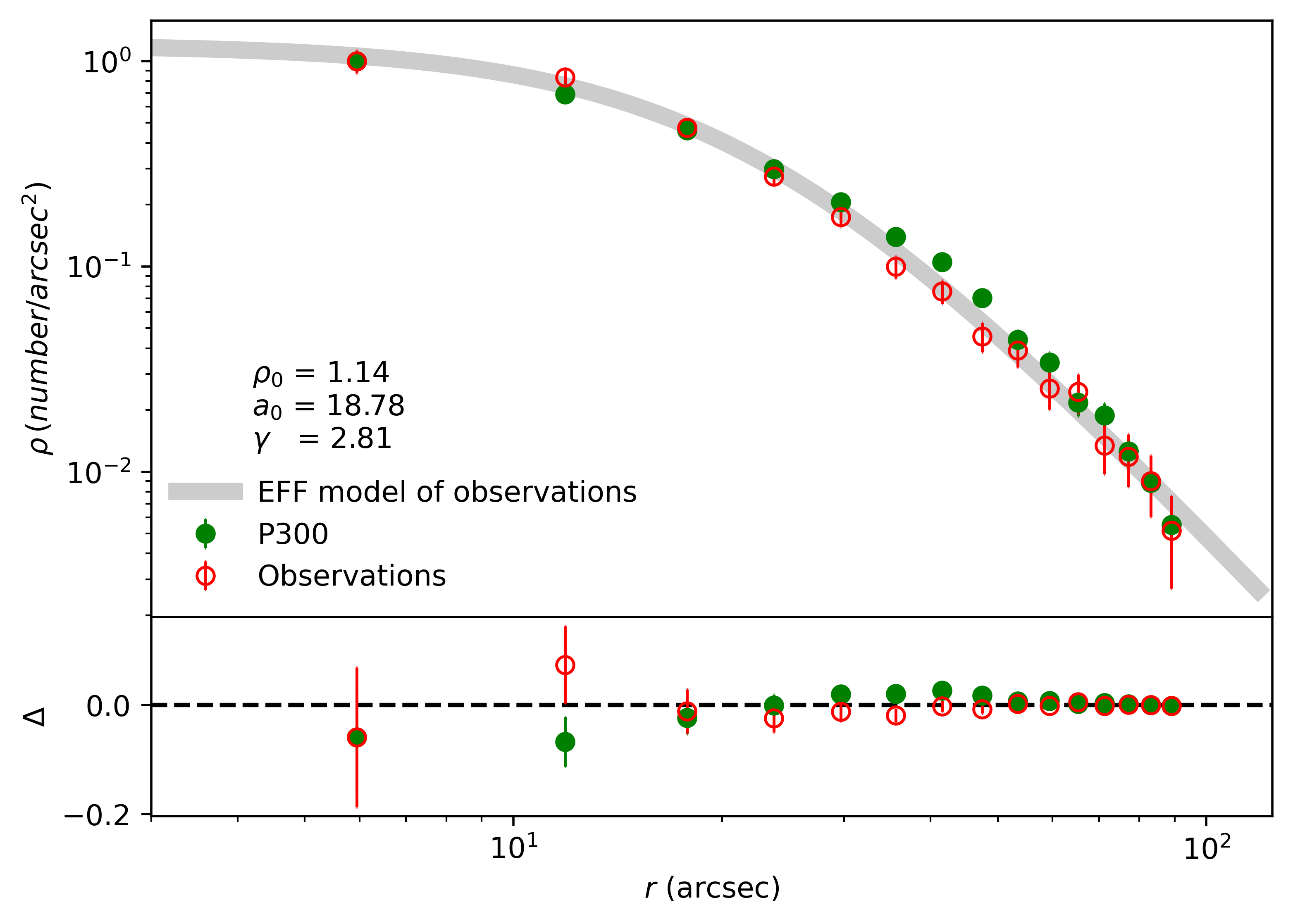}
 
 \caption{The normalized radial number density profile for observations of NGC 1856 (red circles) and best-fitting $N$-body model P300 (green dots). The gray curve is the best-fitting EFF model of the observations of NGC 1856. Both the horizontal and vertical axes are in logarithmic scales. The bottom panel shows the residuals between the best-fit EFF model and density profiles. The error bars for the surface number density profile represent the Poissonian error, $\delta \rho = \rho/\sqrt{N}$, where $N$ is the number of stars in each bin. \label{fig:profile}}
 \end{figure}

 \begin{equation}
 \rho(r)=\rho_0\left[1+\left(\frac{r}{a_0}\right)^2\right]^{-\gamma / 2}
 \end{equation}
 Here, $\rho$ is the number density; $r$ is the distance from a star to the cluster's center; $\rho_0$ is the central number density; $\gamma$ is the power-law slope at large radii; and $a_0$ is the scale radius. The derived number density profile for NGC 1856 (red circles) and the best-fitting EFF model (gray curve) are presented in Figure \ref{fig:profile}. 
We also compute the core radius, $r_{\rm c}$, using Equation \ref{rc} from \citet{1987ApJ...323...54E}:
 \begin{equation}
 r_c \approx a_0 \sqrt{2^{2 / \gamma}-1}
 \label{rc}
 \end{equation}
 {The distance modulus of NGC 1856 is 18.35 mag \citep{10.1093/mnras/stv829}}. NGC 1856 has a core radius of $\sim$ 15.08 arcsec. {As a comparison}, the best-fitting $N$-body model P300 (green dots) is also displayed in Figure \ref{fig:profile}. The core radius of the P300 model is $\sim$ 14.98 arcsec, which is in good agreement with the observations. We only select massive stars because in observation these stars have higher completeness than fainter stars. We have confirmed that the density profile does not dramatically change if we used more faint stars. 

\begin{figure*}[ht!]
  \epsscale{1.1}
 \plottwo{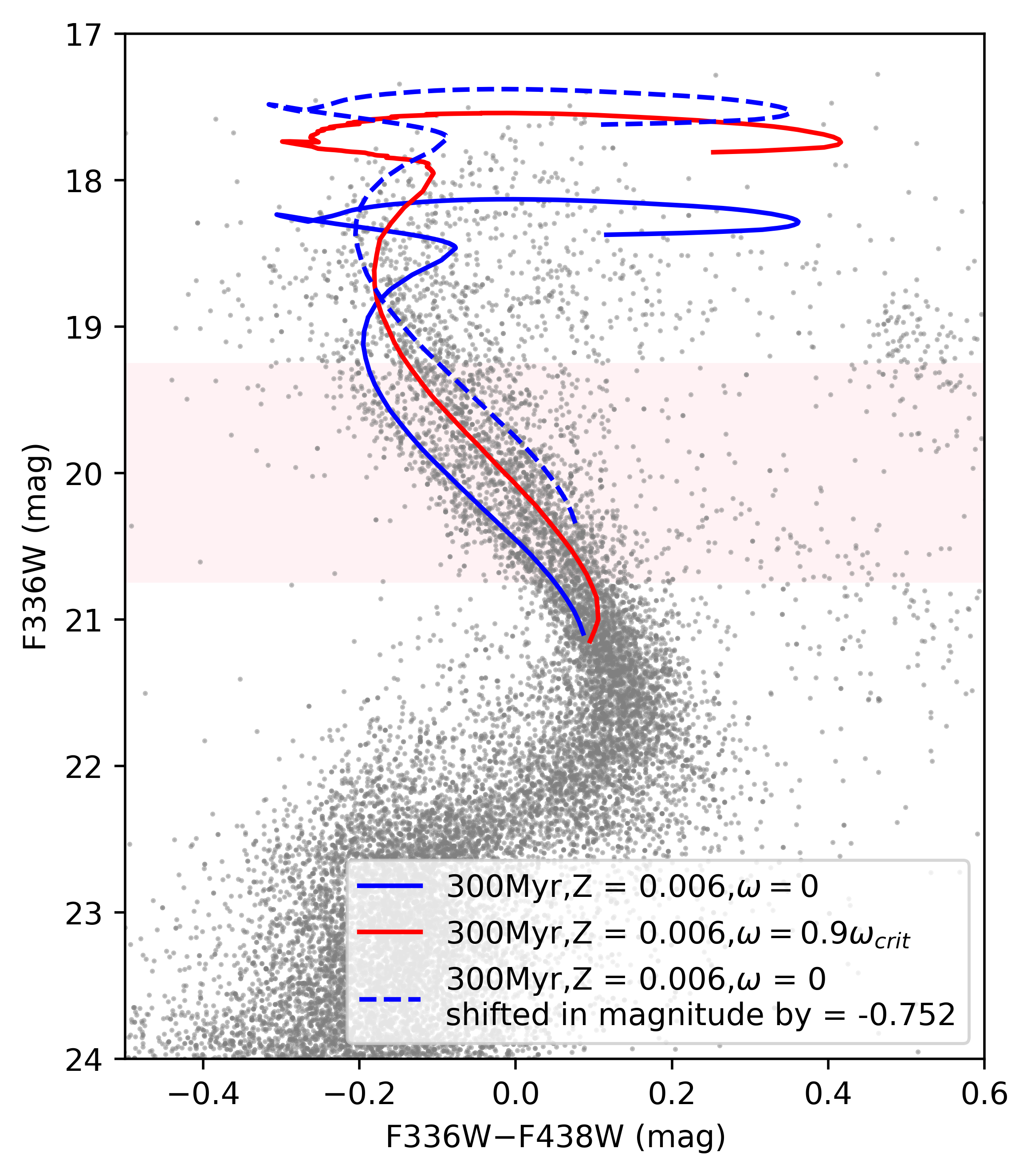}{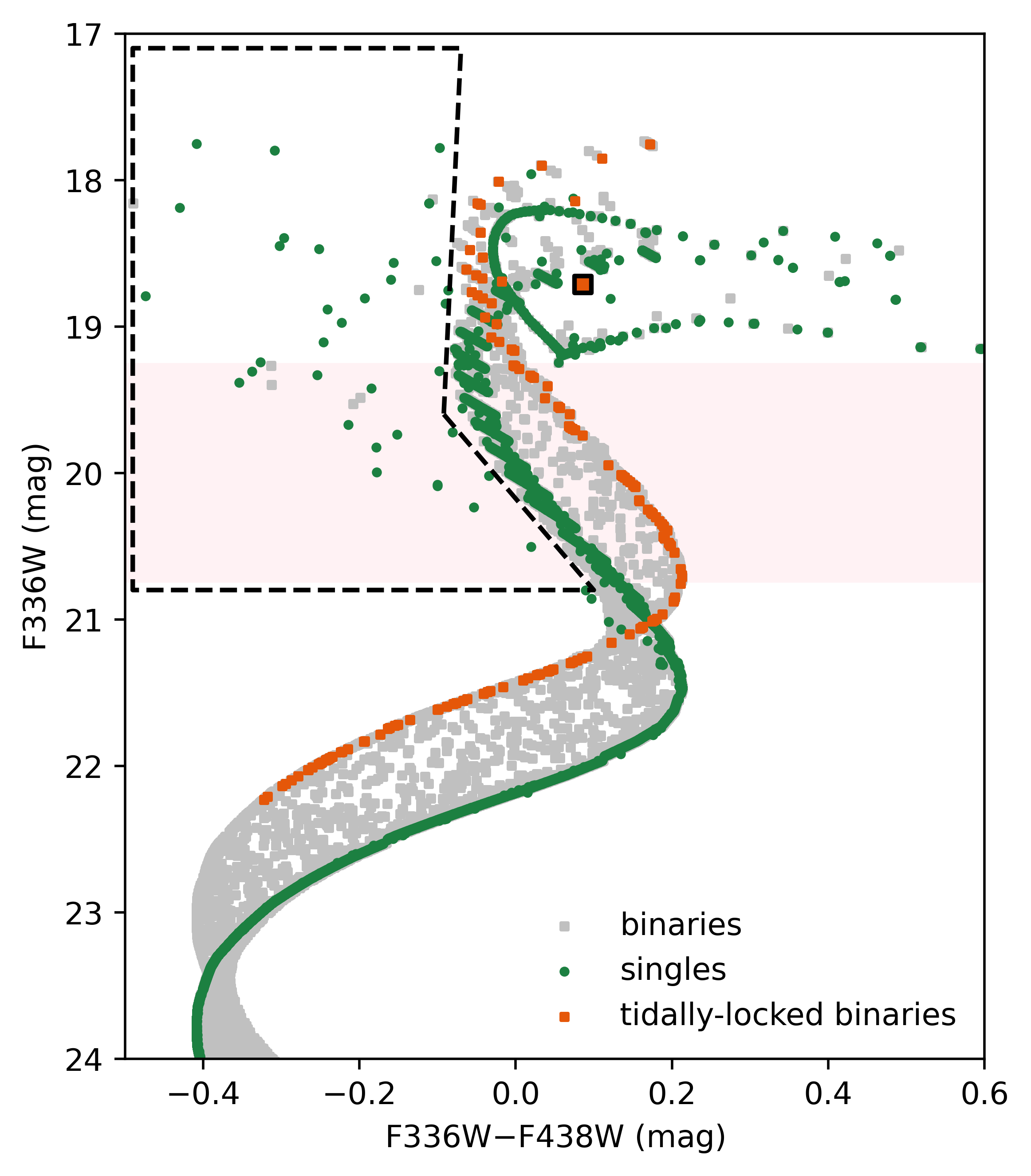}
 \caption{Comparison between the observations of NGC 1856 and $N$-body model P300. (Left) CMD of the two MS components of NGC 1856 and the best-fitting isochrones with the same ages and metallicities but with different rotation rates (solid blue line: nonrotating; solid red line: $\omega = 0.9$ $\omega_{\mathrm{crit}}$), as indicated. The blue dashed line corresponds to the 300 Myr isochrone with $\omega = 0$ shifted in magnitude by -0.752. (Right) The single stars, binary stars, and tidally-locked binary stars in P300 are labeled with green dots, gray squares, and orange squares, respectively. In addition, one orange square in the black box is a dynamically formed tidally-locked binary. The black dashed lines draw the adopted selection box of BSSs. In both panels, stars in the pink background were selected as sample stars to compare their number distributions in Section \ref{sec:results}.}
 \label{fig:cmd}
 \end{figure*}

In Figure \ref{fig:cmd}, we show the CMDs of NGC 1856 and P300, along with the best-fitting isochrones including stellar rotation from the Synthetic CLusters Isochrones \& Stellar Tracks (SYCLIST) \citep{georgy_populations_2013} to the two different MS components in NGC 1856.

\section{Results} \label{sec:results}
\subsection{Tidally-locked Binaries \label{subsec:tb}}

 We trace the evolution of tidally-locked binaries {obtained from the $N$-body simulations} and explore their properties at 300 Myr. Tidal locking is the synchronization of the primary and secondary stellar rotation rates, and can effectively transfer angular momentum from the stellar spin to the orbit \citep{de_Mink_2013}. It can be estimated quantitatively by calculating the synchronization timescales, $\tau_{\mathrm{sync}}$. Following \citet{10.1046/j.1365-8711.2002.05038.x}, we estimate $\tau_{\mathrm{sync}}$ for MS binary stars with radiative envelopes ($M_{\rm p} \geqslant$ 1.25 $M_{\odot}$),
\begin{equation}
 \frac{1}{\tau_{\mathrm{sync}}}=5\times 2^{5 / 3}\left(\frac{G M_{\rm p}}{R_{\rm p}^3}\right)^{1 / 2} \frac{M_{\rm p} R_{\rm p}^2}{{I_{\rm p}}} q^2\left(1+q\right)^{5 / 6} E_2\left(\frac{R_{\rm p}}{a}\right)^{17 / 2}
 \label{eq:tsync}
\end{equation}
Here, $G$ is the gravitational constant; $M_{\rm p}$ and $R_{\rm p}$ are the primary star's mass and radius, respectively; $q$ is the mass ratio; {$I_{\rm p}$ is the moment of inertia of the primary star}; and $E_2$ is a second-order tidal coefficient that can be fitted to values given by \citet{1975A&A....41..329Z}, i.e.,
\begin{equation}
 E_2=1.592 \times 10^{-9} (M_{\rm p}/M_{\odot})^{2.84}
\end{equation}

\begin{figure}[ht!]
 \epsscale{1.18}
 \plotone{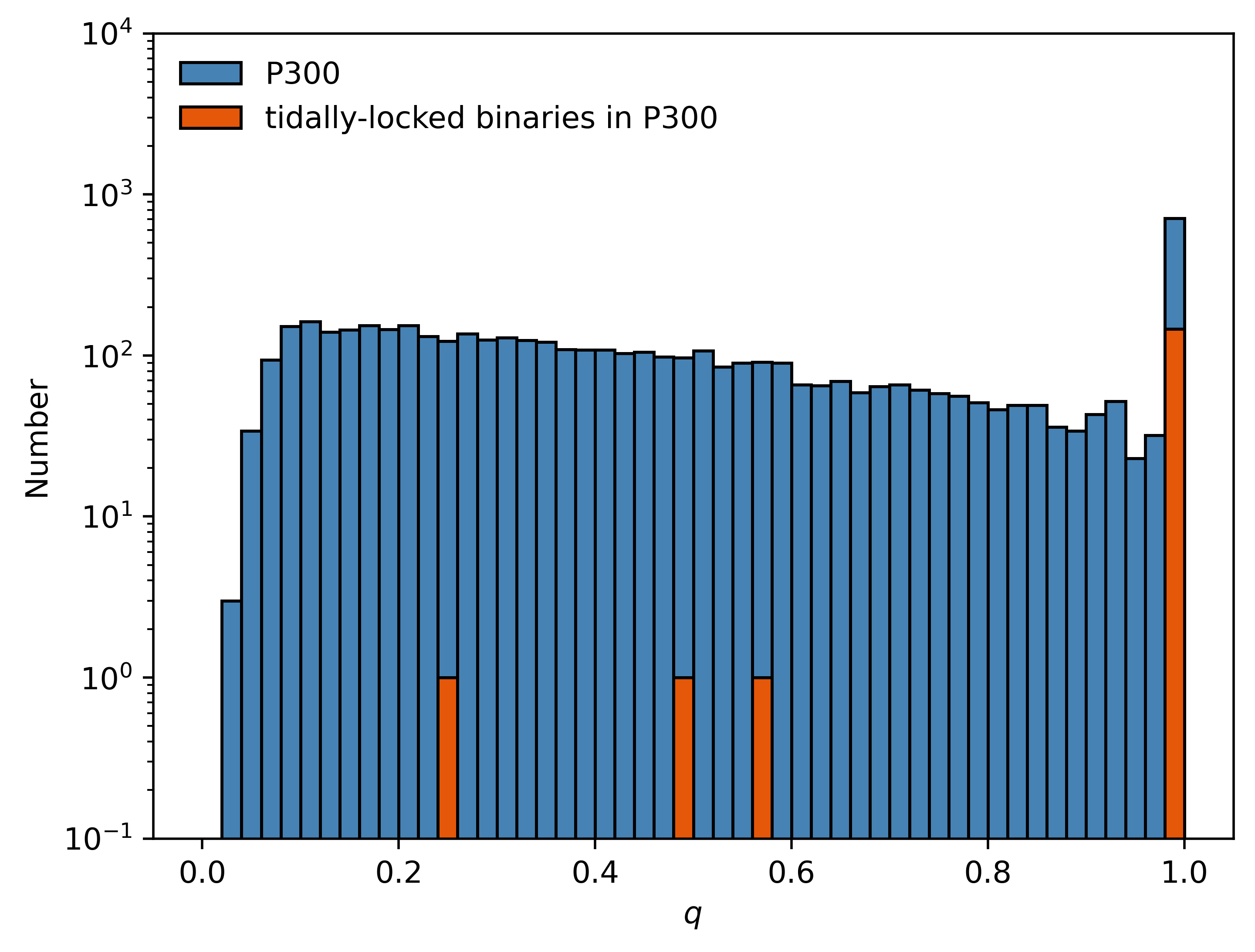}
 \caption{Histogram of the mass-ratio distributions of the MS binaries (blue) and MS tidally-locked binaries (orange). Here only displays MS binary systems with $M_{\rm p} \geqslant$ 1.25 $M_{\odot}$. The vertical axis is on a logarithmic scale. \label{qn}}
\end{figure}

We have confirmed that the components of tidally-locked binaries selected through the Equation \ref{eq:tsync} do exhibit rotational synchronization with the orbit. We finally selected 149 MS binary systems that have been tidally locked in the P300 model. The majority of the resulting tidally-locked binaries are characterized by close separations and circular orbits in the right panel of Figure \ref{fig:ae} corresponding towards an equilibrium state of minimum energy. In Figure \ref{qn}, we find that 146 of them are equal-mass binaries except for three binaries. These tidally-locked binaries account for 21\% of the equal-mass binaries with $M_{\rm p} \geqslant 1.25 M_{\odot}$. According to Equation \ref{eq:binary}, equal-mass binaries will populate a sequence that is brighter than the MS (for single stars) in $\sim$ 0.752 mag. Thus all equal-mass binaries will locate on the red side of the MS. That means these unresolved tidally-locked binaries will be significantly redder than their lower-mass-ratio counterparts. Obviously, the loci of equal-mass tidally-locked binaries in Figure \ref{fig:cmd} are too red to account for the origin of the bMS. 

\begin{figure*}[ht!]
\epsscale{0.95}
 \plotone{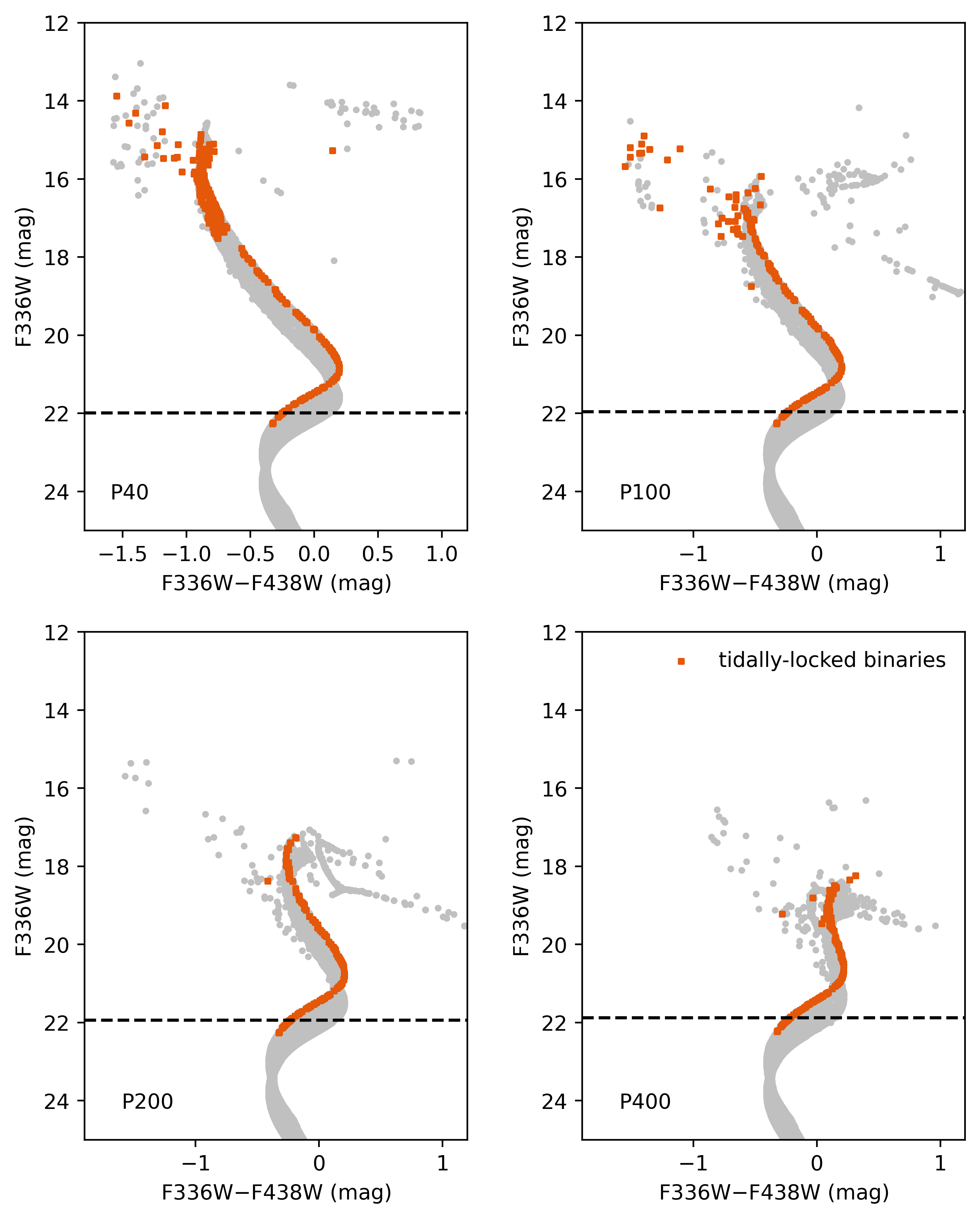}
 \caption{CMDs of P40, P100, P200, P400. 
 The tidally-locked binaries are labeled with orange squares. The black dashed lines correspond to the F336W magnitude of single stars, with a mass of 1.6 $M_{\odot}$. \label{fig:CMDother}}
 \end{figure*}

Only one tidally-locked binary with $q = 0.49$ is dynamically formed, while the progenitors of other tidally-locked binaries are primordial binaries. Tidally-locked binary systems, produced through dynamic encounters in dense clusters, are much less efficient than typical stellar evolutionary processes. In Figure \ref{fig:CMDother}, we compare the CMDs of tidally locked binaries at different ages: 40 Myr, 100 Myr, 200 Myr, and 400 Myr. These ages roughly correspond to the ages of the NGC 1818 \citep{cordoni_ngc1818_2022}, and NGC 1850 \citep{milone_hubble_2022}, NGC 1866 \citep{dupree_ngc_2017}, NGC 3532 \citep{cordoni_extended_2018}, respectively. These clusters all exhibit significant split MSs or an eMSTO. We find that at the very early stage (models earlier than P40), there are some low-mass-ratio tidally-locked binaries. However, these binaries all have primary stars heavier than 5.0 $M_{\odot}$. That means they can only populate the blue side of the upper MS. 
For the lower part of the MS (with F336W $\lesssim$ 18.0 mag, corresponding to B9 stars at the distance of the LMC), only equal-mass binaries can be tidally-locked. As a result, they only populate the red side of the MS. Therefore for clusters older than $\sim$ 100 Myr, as illustrated in Figure \ref{fig:CMDother}, no tidally-locked binaries can be found in the blue side of the MS.

To facilitate a direct comparison of the work in \citetalias[]{Li_2017}, we select two subpopulations: the tidally-locked binary subpopulation and single star subpopulation with 18.25 mag $\leq$ F336W $\leq$ 19.75 mag (stars in the pink background of the right panel of Figure \ref{fig:cmd}). The resulting number ratio of stars associated with both subpopulations is $N_{\mathrm{tidal}}/N_{\mathrm{single}}$ = 2\%/98\%. The single stars significantly dominate the stellar numbers. However, the resulting number ratio of stars associated with both bMS and rMS in \citetalias[]{Li_2017} is $N_{\mathrm{bMS}}/N_{\mathrm{rMS}}=34\%/66\%$.

\begin{figure}[ht!]
  \epsscale{1.18}
  \plotone{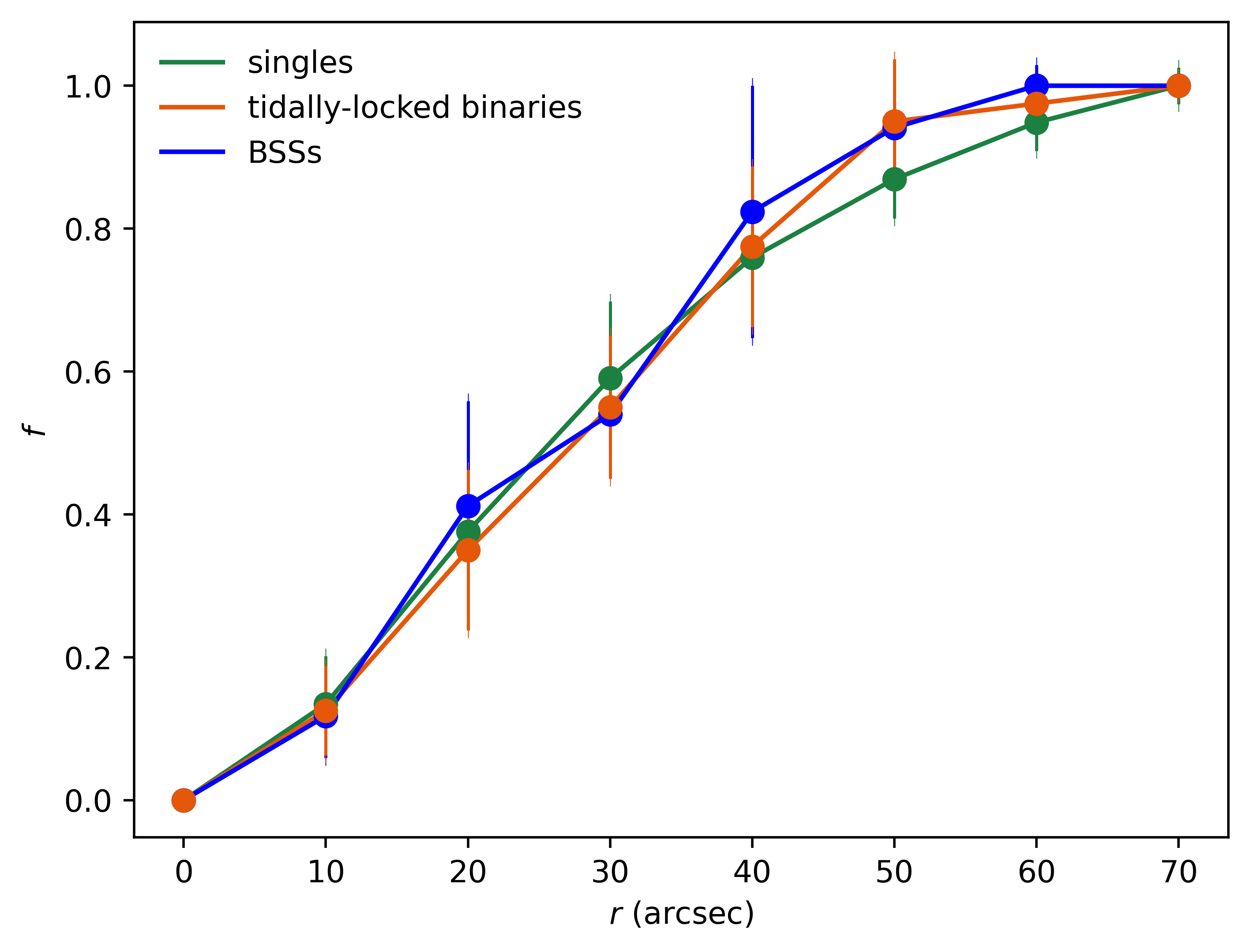}
  \caption{Cumulative number distributions of three stellar subpopulations in P300 with 18.25 mag $\leq$ F336W $\leq$ 19.75 mag: singles (green), tidally-locked binaries (orange) and BSSs (blue). \label{fig:number}}
 \end{figure}

 \citetalias[]{Li_2017} analyzed the number ratio versus radius of the stars belonging to each split MS component in NGC 1856 (stars in the pink background of the left panel of Figure \ref{fig:cmd}). They found that their number ratios remain approximately constant from the central region to the cluster's periphery (about five times the core radius, see Figure 5 in \citetalias[]{Li_2017}), suggesting that the two components are homogeneously distributed in space. Next, we similarly examine whether the tidally-locked binary subpopulation exhibits a different spatial distribution from single stars, by inspecting cumulative profiles of their number distributions in Figure \ref{fig:number}. To study the radial distributions of the selected subpopulations, we divide the subpopulations into seven concentric circles with radii spanning from 10-70 arcsec, like \citetalias[]{Li_2017}, and count the number of stars in each circle. We find that there seems to be no difference in the spatial distributions of these two subpopulations within the error range.

In summary, although our results report a consistency between the radial distribution of the tidally-locked binaries and the observed bMS stars in \citetalias[]{Li_2017}, these tidally-locked binaries will not correctly populate a bMS given that they are all equal-mass binaries and cannot be resolved by the HST. Counterparts with low-mass ratios could not constitute a statistical effect due to their rarity. The resulting number ratio $N_{\mathrm{tidal}}/N_{\mathrm{single}}$ is too low compared with the observation.

\subsection{Blue straggler stars \label{subsec:bss}}

BSSs are massive MS stars that are much younger than the bulk of cluster member stars. They are produced by mass transfer or direct merger of two binary components. Young clusters evolve very rapidly, they will produce a significant difference between the zero-age MS and the present-age MS. Thus, the position of the BSSs (refer to the blue binary-evolutionary-induced stars) in young clusters on the CMD is the same as that of the bMS, and even BSSs below the MSTO fall in a bluer region than the MS as displayed in Figure \ref{fig:cmd} (stars in the black dashed box).

We use the same selection criteria as described in Section \ref{subsec:tb} (18.25 mag $\leq$ F336W $\leq$ 19.75 mag) to obtain the BSS subpopulation and compare it with single star subpopulation, generating a number ratio associated with both subpopulations of $N_{\mathrm{BSS}}/N_{\mathrm{single}}$ = 1\%/99\% (only 19 BSSs). The total number of the BSS subpopulation is half of the tidally-locked binary subpopulation. 
A significant fraction of the observed bMS is hard to interpret because of the scarcity of BSSs. 
In Figure \ref{fig:number}, we find that the cumulative number distributions of two subpopulations, namely BSSs and single stars, successfully reproduce the results of \citetalias[]{Li_2017}. This result indicates that at the age of the NGC 1856, BSSs are fully mixed with single stars. 


\section{Discussions} \label{sec:discussions}

Primordial mass segregation and tidal field were not included in our simulation. The question of whether mass segregation is likely to be a primordial outcome of star formation (e.g., \citealt{bonnell_mass_1998}), whether it is dynamical (e.g., \citealt{10.1093/mnras/stt2231}), or a combination of the two \citep{10.1111/j.1365-2966.2009.15499.x} has been debated without definitive explanations. If primordial mass segregation occurs, we expect that both the tidally-locked binaries and BSSs will form a more compact configuration than single stars, in that case they cannot account for the observed spatial distributions of the bMS and rMS. We do not take the Galactic potential into account in our model, because NGC 1856 is located in the LMC, about $\sim$ 50 kpc from the Galactic center, we expect that the tidal field from the Galaxy is negligible. Although NGC 1856 may be affected by the LMC, the influence of the LMC tidal field is unknown, as only the projected distance between the cluster and the LMC center is known. 

In figure \ref{fig:ae}, we find that the tidally-locked binary systems at 300 Myr have short orbital periods. According to Equation \ref{eq:tsync}, the synchronization timescale is highly dependent on the binary semi-major axis. Binary systems with long orbital periods, thus large semi-major axis, would have much weaker tidal torques than close binary systems. The long-term dynamical evolution of star clusters would disrupt long-period binaries, with negligible effect on tidally-locked binaries with short periods. Therefore almost all tidally-locked binaries are evolved from primordial binaries. Black hole (BH) systems also affect the dynamical evolution of a star cluster. \citet{wang_impact_2022} found that massive BH binaries dominate the binary heating process by strongly suppressing the dynamical impact on the low-mass binaries. At 300 Myr, most BHs have not escaped from the cluster. The dynamically formed binaries tend to have long orbital periods and low mass ratios. Due to the rigor of tidal locking and the suppressed formation of low-mass binaries, dynamically formed binaries have almost no chance to contribute to the tidally-locked binaries with low-mass ratios. 

In Figure \ref{fig:CMDother}, we present four CMDs, which demonstrates the evolution of tidally-locked binaries in CMDs. On timescales shorter than 100 Myr, there are tidally-locked binaries with small mass ratios in the simulated cluster, but these binaries all have their primary stars more massive than {5.0 $M_{\odot}$}. These tidally-locked binaries possibly have blue colors because their fluxes are dominated by primary, slowly rotating stars. Taking 40 Myr as an example, massive binaries have shorter synchronization timescales than their less massive counterparts. According to Equation \ref{eq:tsync}, the synchronization timescales depend on both the primary star's mass and the mass ratio. Only a binary system with a massive primary star can be rapidly tidally locked if its mass ratio is small. For binaries with less massive primary stars, a higher mass ratio is required to make the system being rapidly synchronized. As a result, low mass-ratio tidally-locked binaries can only populate a blue sequence to upper MS. For clusters older than $\sim$ 100 Myr, these low mass-ratio tidally-locked binaries will have their primary stars evolved off. The remaining tidally-locked binaries will be all high mass-ratio (close to unique) binaries, located on the red side of the MS. In summary, our simulation indicates that tidally-locked binaries cannot account for the observed bMS of NGC 1856, and this may also apply to other clusters with different ages, as observations clearly exhibit a split pattern down to the F-type MS range \citep[e.g.,][]{10.1093/mnras/sty661}. 

The number of tidally-locked binaries at $\sim$ 300 Myr depends on both the IMF and the initial binary period distribution. Under such assumptions, the number of tidally-locked binaries is quite sparse. In our simulation, a canonic Kroupa-like IMF is applied \citep{10.1046/j.1365-8711.2001.04022.x}, although a recent study suggests that the stellar IMF may vary with metallicity and formation time \citep{li_stellar_2023}, which would not change our results dramatically as long as the IMF remains bottom-heavy. According to \citetalias[]{Li_2017}, the observed number of bMS stars in NGC 1856 is over $\sim$ 1000. As a comparison, in our simulation, there are only 50 tidally-locked binaries; among them, only three are low mass-ratio binaries formed through dynamics. 

Our simulation indicates that the tidally-locked binaries are fully mixed with single stars in space. This result is in contrast to \cite{yang_spatial_2021}, in their simulations tidally-locked binaries are highly mass segregated. This difference may be due to a longer relaxation time of NGC 1856 than clusters in \cite{yang_spatial_2021}. To examine the dynamical age of P300, we use the formula introduced by \citet{1987degc.book.....S} to evaluate its current relaxation time,
\begin{equation}
 T_{\mathrm{rh}}=0.138 \frac{N^{1 / 2} R_{\mathrm{h}}^{3 / 2}}{\langle m\rangle^{1 / 2} G^{1 / 2} \ln \Lambda}
 \label{eq:trh}
\end{equation}
where $N$ is the number of stars excluding BHs in P300, which is 265,902 with a total mass of 1.4$\times 10^5 M_{\odot}$. $R_{\mathrm{h}}$ is the half-mass radius equal to 2.93 pc. $\langle m\rangle$ is the average mass ($\sim 0.53$ $M_{\odot}$ in our simulation), and the Coulomb logarithm ln$\Lambda$ is $0.02N$ based on the measurement of \citet{1996MNRAS.279.1037G}. The resulting half-mass relaxation time is $\sim$ 852.3 Myr, therefore the dynamical age of NGC 1856 is only one-third of its relaxation time. The simulated cluster at 300 Myr is still under relaxing. As a comparison, the simulation made by \cite{yang_spatial_2021} assumes 100,000 particles and a half-mass radius of $R_{\rm h} = 0.5$ pc. If the average mass of their simulated cluster is comparable to ours, we estimate that their half-mass relaxation time would be $\sim$ 100 Myr. 
 
The mass-segregation timescale, $T_{\mathrm{ms}}$, for specific stellar populations is defined as \citealt{1987degc.book.....S}:
\begin{equation}
  T_{\mathrm{ms}}=\frac{\langle m\rangle}{m_1} T_{\mathrm{rh}}
\end{equation}
where $ m_1$ is the average mass of a specific stellar population ($\sim2.85$ $M_{\odot}$ for BSSs and $\sim4.15$ $M_{\odot}$ for tidally-locked binaries in our simulation). The resulting mass-segregation timescales are $\sim$124 Myr and $\sim$109 Myr, respectively. \citet{wang_survival_2020} found that the BHs achieve mass segregation after at least five times the mass-segregation timescale of BHs, i.e., $5T_{\rm ms}$. However, this timescale is older than the current age of the cluster. \citet{baumgardt_global_2017} found a decrease in the amount of mass segregation or complete suppression of mass segregation due to stellar-mass BHs. 
 Thus the tidally-locked binaries and BSSs in Figure \ref{fig:number} show similar spatial distributions with normal stars. This is why in the simulation of \cite{yang_spatial_2021}, their tidally-locked binaries are fully mass segregated while our simulation does not. We note that the above analysis should be treated with caution. The relaxation timescales in Equation \ref{eq:trh} vary with the total mass and density of the cluster and significantly change from one cluster to another. Our simulations can only interpret the mass segregation of their stellar components in an NGC 1856-like cluster. 

Although the positions of BSSs do occupy the bMS region, in our simulation, we only produce 19 BSSs through binary interaction. This number cannot account for the observation. To reproduce the observed number ratio of \citetalias[]{Li_2017} ($N_{\mathrm{bMS}}/N_{\mathrm{rMS}}=34\%/66\%$), about $\sim$1200 BSSs are required. The BSSs produced in $N$-body models through the classical binary evolution alone cannot explain the observed number ratio in \citetalias[]{Li_2017}. If BSSs are responsible for the observed bMS, a lot of BSSs may be produced during the early formation stage of the cluster, like what was suggested by \citet{wang2022stellar}. A large fraction of BSSs might be generated by stellar mergers that occurred during the PMS phase, through tidal interactions with their circumstellar material \citep[e.g.,][]{korntreff_towards_2012}.

To examine the contribution of binary interactions, one method is to investigate if the bMS stars have a more significant radial velocity (RV) dispersion than normal stars, as close binary systems are usually RV variables. We have briefly studied this effect using our simulated tidally-locked binaries. If we adopt the straightforward assumption that the distribution of stellar rotation axes is stochastic, we find that half of our tidally-locked binaries would have RV variations $>$ 50 $\rm km s^{-1}$ in two-epoch velocity measurements over a timescale of $\sim$ 1 yr. 
If we assume that the stellar-spin axes are aligned ($i$ = $45^{\circ}$), half of them are expected to have RV variations $>$ 95 $\rm km s^{-1}$. These predictions can be tested immediately for many Galactic open clusters through time-domain spectroscopic surveys.

\section{Conclusions} \label{sec:conclusions}

This research aims to verify whether the MS tidally-locked binaries and BSSs account for the observed split MSs in an NGC 1856-like cluster by using {\tt\string PETAR} code. Our main results and conclusions are summarized as below:

\begin{enumerate}[label=(\arabic*)]
\item Both tidally-locked binaries and BSSs are fully mixed with single stars in space at the age of NGC 1856, which is similar to the spatial distributions of bMS and rMS in observation. 
 \item However, tidally-locked binaries are almost all equal-mass binaries at the age of NGC 1856. These unresolved binaries will populate a redder sequence to the MS, and thus cannot explain the observed bMS. Tidally-locked binaries cannot explain the bMS observed in younger clusters because they cannot populate the lower part of the bMS at any time. The number ratio between tidally-locked binaries and single stars is $N_{\mathrm{tidal}}/N_{\mathrm{single}}$ = 2\%/98\%, which is dramatically different from the observation of \citetalias[]{Li_2017}. 
 \item Although BSSs do populate a blue sequence to the MS in the CMD, the number of BSSs cannot explain the observation neither, at least for BSSs generated by classical binary evolution. Lots of primordial BSSs formed through binary mergers during the very early-stage of the cluster are required, probably through tidally interactions with their circumstellar discs, as suggested by \citet{wang2022stellar}. 
\end{enumerate}

\begin{acknowledgements}
{C. L. is supported by the National Key R\&D Program of China (2020YFC2201400). This work was supported by the National Natural Science Foundation of China through grant 12233013. }
\end{acknowledgements}

\vspace{5mm}

\software{{\tt\string Astropy} \citep{2013A&A...558A..33A,2018AJ....156..123A},   
{\tt\string Matplotlib} \citep{4160265},
{\tt\string SciPy} \citep{Scipy},
{\tt\string NumPy} \citep{numpy},
{\tt\string SYCLIST} \citep{georgy_populations_2013},
{\tt\string PETAR} \citep{10.1093/mnras/staa1915},
{{\tt\string MCLUSTER} \citep{10.1111/j.1365-2966.2011.19412.x,10.1093/mnras/sty2232}},
{\tt\string YBC} \citep{YBC},
          }

\bibliography{NGC1856}{}

\begin{thebibliography}{}
\expandafter\ifx\csname natexlab\endcsname\relax\def\natexlab#1{#1}\fi
\providecommand{\url}[1]{\href{#1}{#1}}
\providecommand{\dodoi}[1]{doi:~\href{http://doi.org/#1}{\nolinkurl{#1}}}
\providecommand{\doeprint}[1]{\href{http://ascl.net/#1}{\nolinkurl{http://ascl.net/#1}}}
\providecommand{\doarXiv}[1]{\href{https://arxiv.org/abs/#1}{\nolinkurl{https://arxiv.org/abs/#1}}}

\bibitem[{{Aarseth} {et~al.}(1974){Aarseth}, {Henon}, \&
  {Wielen}}]{1974A&A....37..183A}
{Aarseth}, S.~J., {Henon}, M., \& {Wielen}, R. 1974, \aap, 37, 183

\bibitem[{Andronov {et~al.}(2006)Andronov, Pinsonneault, \&
  Terndrup}]{andronov_mergers_2006}
Andronov, N., Pinsonneault, M.~H., \& Terndrup, D.~M. 2006, The Astrophysical
  Journal, 646, 1160, \dodoi{10.1086/505127}

\bibitem[{{Astropy Collaboration} {et~al.}(2013){Astropy Collaboration},
  {Robitaille}, {Tollerud}, {Greenfield}, {Droettboom}, {Bray}, {Aldcroft},
  {Davis}, {Ginsburg}, {Price-Whelan}, {Kerzendorf}, {Conley}, {Crighton},
  {Barbary}, {Muna}, {Ferguson}, {Grollier}, {Parikh}, {Nair}, {Unther},
  {Deil}, {Woillez}, {Conseil}, {Kramer}, {Turner}, {Singer}, {Fox}, {Weaver},
  {Zabalza}, {Edwards}, {Azalee Bostroem}, {Burke}, {Casey}, {Crawford},
  {Dencheva}, {Ely}, {Jenness}, {Labrie}, {Lim}, {Pierfederici}, {Pontzen},
  {Ptak}, {Refsdal}, {Servillat}, \& {Streicher}}]{2013A&A...558A..33A}
{Astropy Collaboration}, {Robitaille}, T.~P., {Tollerud}, E.~J., {et~al.} 2013,
  \aap, 558, A33, \dodoi{10.1051/0004-6361/201322068}

\bibitem[{{Astropy Collaboration} {et~al.}(2018){Astropy Collaboration},
  {Price-Whelan}, {Sip{\H{o}}cz}, {G{\"u}nther}, {Lim}, {Crawford}, {Conseil},
  {Shupe}, {Craig}, {Dencheva}, {Ginsburg}, {VanderPlas}, {Bradley},
  {P{\'e}rez-Su{\'a}rez}, {de Val-Borro}, {Aldcroft}, {Cruz}, {Robitaille},
  {Tollerud}, {Ardelean}, {Babej}, {Bach}, {Bachetti}, {Bakanov}, {Bamford},
  {Barentsen}, {Barmby}, {Baumbach}, {Berry}, {Biscani}, {Boquien}, {Bostroem},
  {Bouma}, {Brammer}, {Bray}, {Breytenbach}, {Buddelmeijer}, {Burke},
  {Calderone}, {Cano Rodr{\'\i}guez}, {Cara}, {Cardoso}, {Cheedella}, {Copin},
  {Corrales}, {Crichton}, {D'Avella}, {Deil}, {Depagne}, {Dietrich}, {Donath},
  {Droettboom}, {Earl}, {Erben}, {Fabbro}, {Ferreira}, {Finethy}, {Fox},
  {Garrison}, {Gibbons}, {Goldstein}, {Gommers}, {Greco}, {Greenfield},
  {Groener}, {Grollier}, {Hagen}, {Hirst}, {Homeier}, {Horton}, {Hosseinzadeh},
  {Hu}, {Hunkeler}, {Ivezi{\'c}}, {Jain}, {Jenness}, {Kanarek}, {Kendrew},
  {Kern}, {Kerzendorf}, {Khvalko}, {King}, {Kirkby}, {Kulkarni}, {Kumar},
  {Lee}, {Lenz}, {Littlefair}, {Ma}, {Macleod}, {Mastropietro}, {McCully},
  {Montagnac}, {Morris}, {Mueller}, {Mumford}, {Muna}, {Murphy}, {Nelson},
  {Nguyen}, {Ninan}, {N{\"o}the}, {Ogaz}, {Oh}, {Parejko}, {Parley}, {Pascual},
  {Patil}, {Patil}, {Plunkett}, {Prochaska}, {Rastogi}, {Reddy Janga},
  {Sabater}, {Sakurikar}, {Seifert}, {Sherbert}, {Sherwood-Taylor}, {Shih},
  {Sick}, {Silbiger}, {Singanamalla}, {Singer}, {Sladen}, {Sooley},
  {Sornarajah}, {Streicher}, {Teuben}, {Thomas}, {Tremblay}, {Turner},
  {Terr{\'o}n}, {van Kerkwijk}, {de la Vega}, {Watkins}, {Weaver}, {Whitmore},
  {Woillez}, {Zabalza}, \& {Astropy Contributors}}]{2018AJ....156..123A}
{Astropy Collaboration}, {Price-Whelan}, A.~M., {Sip{\H{o}}cz}, B.~M., {et~al.}
  2018, \aj, 156, 123, \dodoi{10.3847/1538-3881/aabc4f}

\bibitem[{Banerjee {et~al.}(2020)Banerjee, Belczynski, Fryer, Berczik, Hurley,
  Spurzem, \& Wang}]{refId1}
Banerjee, S., Belczynski, K., Fryer, C.~L., {et~al.} 2020, A\&A, 639, A41,
  \dodoi{10.1051/0004-6361/201935332}

\bibitem[{Bastian \& de~Mink(2009)}]{bastian_effect_2009}
Bastian, N., \& de~Mink, S.~E. 2009, Monthly Notices of the Royal Astronomical
  Society: Letters, 398, L11, \dodoi{10.1111/j.1745-3933.2009.00696.x}

\bibitem[{Bastian {et~al.}(2020)Bastian, Kamann, Amard, Charbonnel, Haemmerlé,
  \& Matt}]{10.1093/mnras/staa1332}
Bastian, N., Kamann, S., Amard, L., {et~al.} 2020, Monthly Notices of the Royal
  Astronomical Society, 495, 1978, \dodoi{10.1093/mnras/staa1332}

\bibitem[{Bastian {et~al.}(2018)Bastian, Kamann, Cabrera-Ziri, Georgy,
  Ekström, Charbonnel, de Juan Ovelar, \& Usher}]{bastian_extended_2018}
Bastian, N., Kamann, S., Cabrera-Ziri, I., {et~al.} 2018, Monthly Notices of
  the Royal Astronomical Society, 480, 3739, \dodoi{10.1093/mnras/sty2100}

\bibitem[{Bastian \& Niederhofer(2015)}]{bastian_morphology_2015}
Bastian, N., \& Niederhofer, F. 2015, Monthly Notices of the Royal Astronomical
  Society, 448, 1863, \dodoi{10.1093/mnras/stv116}

\bibitem[{Bastian \& Strader(2014)}]{bastian_constraining_2014}
Bastian, N., \& Strader, J. 2014, Monthly Notices of the Royal Astronomical
  Society, 443, 3594, \dodoi{10.1093/mnras/stu1407}

\bibitem[{Bastian {et~al.}(2017)Bastian, Cabrera-Ziri, Niederhofer, de~Mink,
  Georgy, Baade, Correnti, Usher, \& Romaniello}]{bastian_high_2017}
Bastian, N., Cabrera-Ziri, I., Niederhofer, F., {et~al.} 2017, Monthly Notices
  of the Royal Astronomical Society, 465, 4795, \dodoi{10.1093/mnras/stw3042}

\bibitem[{Baumgardt \& Sollima(2017)}]{baumgardt_global_2017}
Baumgardt, H., \& Sollima, S. 2017, Monthly Notices of the Royal Astronomical
  Society, 472, 744, \dodoi{10.1093/mnras/stx2036}

\bibitem[{Belczynski {et~al.}(2010)Belczynski, Bulik, Fryer, Ruiter, Valsecchi,
  Vink, \& Hurley}]{Belczynski_2010}
Belczynski, K., Bulik, T., Fryer, C.~L., {et~al.} 2010, The Astrophysical
  Journal, 714, 1217, \dodoi{10.1088/0004-637X/714/2/1217}

\bibitem[{Belczynski {et~al.}(2016)Belczynski, Holz, Bulik, \&
  O'Shaughnessy}]{Belczynski2016}
Belczynski, K., Holz, D.~E., Bulik, T., \& O'Shaughnessy, R. 2016, Nature, 534,
  512, \dodoi{10.1038/nature18322}

\bibitem[{Belloni {et~al.}(2017)Belloni, Askar, Giersz, Kroupa, \&
  Rocha-Pinto}]{belloni_initial_2017}
Belloni, D., Askar, A., Giersz, M., Kroupa, P., \& Rocha-Pinto, H.~J. 2017,
  Monthly Notices of the Royal Astronomical Society, 471, 2812,
  \dodoi{10.1093/mnras/stx1763}

\bibitem[{Bonnell \& Davies(1998)}]{bonnell_mass_1998}
Bonnell, I.~A., \& Davies, M.~B. 1998, Monthly Notices of the Royal
  Astronomical Society, 295, 691, \dodoi{10.1046/j.1365-8711.1998.01372.x}

\bibitem[{Chen {et~al.}(2019)Chen, Girardi, Fu, Bressan, Aringer, Dal~Tio,
  Pastorelli, Marigo, Costa, \& Zhang}]{YBC}
Chen, Y., Girardi, L., Fu, X., {et~al.} 2019, A\&A, 632, A105,
  \dodoi{10.1051/0004-6361/201936612}

\bibitem[{Cordoni {et~al.}(2018)Cordoni, Milone, Marino, Criscienzo,
  D’Antona, Dotter, Lagioia, \& Tailo}]{cordoni_extended_2018}
Cordoni, G., Milone, A.~P., Marino, A.~F., {et~al.} 2018, The Astrophysical
  Journal, 869, 139, \dodoi{10.3847/1538-4357/aaedc1}

\bibitem[{Cordoni {et~al.}(2022)Cordoni, Milone, Marino, Cignoni, Lagioia,
  Tailo, Carlos, Dondoglio, Jang, Mohandasan, \&
  Legnardi}]{cordoni_ngc1818_2022}
---. 2022, Nature Communications, 13, 4325, \dodoi{10.1038/s41467-022-31977-y}

\bibitem[{Correnti {et~al.}(2017)Correnti, Goudfrooij, Bellini, Kalirai, \&
  Puzia}]{10.1093/mnras/stx010}
Correnti, M., Goudfrooij, P., Bellini, A., Kalirai, J.~S., \& Puzia, T.~H.
  2017, Monthly Notices of the Royal Astronomical Society, 467, 3628,
  \dodoi{10.1093/mnras/stx010}

\bibitem[{Costa {et~al.}(2019)Costa, Girardi, Bressan, Chen, Goudfrooij,
  Marigo, Rodrigues, \& Lanza}]{costa_multiple_2019}
Costa, G., Girardi, L., Bressan, A., {et~al.} 2019, Astronomy \& Astrophysics,
  631, A128, \dodoi{10.1051/0004-6361/201936409}

\bibitem[{D'Antona {et~al.}(2015)D'Antona, Di~Criscienzo, Decressin, Milone,
  Vesperini, \& Ventura}]{dantona_extended_2015}
D'Antona, F., Di~Criscienzo, M., Decressin, T., {et~al.} 2015, Monthly Notices
  of the Royal Astronomical Society, 453, 2638, \dodoi{10.1093/mnras/stv1794}

\bibitem[{D'Antona {et~al.}(2017)D'Antona, Milone, Tailo, Ventura, Vesperini,
  \& Di~Criscienzo}]{d2017stars}
D'Antona, F., Milone, A.~P., Tailo, M., {et~al.} 2017, Nature Astronomy, 1,
  0186, \dodoi{10.1038/s41550-017-0186}

\bibitem[{de~Mink {et~al.}(2013)de~Mink, Langer, Izzard, Sana, \&
  de~Koter}]{de_Mink_2013}
de~Mink, S.~E., Langer, N., Izzard, R.~G., Sana, H., \& de~Koter, A. 2013, The
  Astrophysical Journal, 764, 166, \dodoi{10.1088/0004-637X/764/2/166}

\bibitem[{Dupree {et~al.}(2017)Dupree, Dotter, Johnson, Marino, Milone, Bailey,
  Crane, Mateo, \& Olszewski}]{dupree_ngc_2017}
Dupree, A.~K., Dotter, A., Johnson, C.~I., {et~al.} 2017, The Astrophysical
  Journal Letters, 846, L1, \dodoi{10.3847/2041-8213/aa85dd}

\bibitem[{{Elson} {et~al.}(1987){Elson}, {Fall}, \&
  {Freeman}}]{1987ApJ...323...54E}
{Elson}, R. A.~W., {Fall}, S.~M., \& {Freeman}, K.~C. 1987, \apj, 323, 54,
  \dodoi{10.1086/165807}

\bibitem[{Fryer {et~al.}(2012)Fryer, Belczynski, Wiktorowicz, Dominik,
  Kalogera, \& Holz}]{Fryer_2012}
Fryer, C.~L., Belczynski, K., Wiktorowicz, G., {et~al.} 2012, The Astrophysical
  Journal, 749, 91, \dodoi{10.1088/0004-637X/749/1/91}

\bibitem[{Georgy {et~al.}(2013)Georgy, Ekström, Granada, Meynet, Mowlavi,
  Eggenberger, \& Maeder}]{georgy_populations_2013}
Georgy, C., Ekström, S., Granada, A., {et~al.} 2013, Astronomy \&
  Astrophysics, 553, A24, \dodoi{10.1051/0004-6361/201220558}

\bibitem[{{Giersz} \& {Heggie}(1996)}]{1996MNRAS.279.1037G}
{Giersz}, M., \& {Heggie}, D.~C. 1996, \mnras, 279, 1037,
  \dodoi{10.1093/mnras/279.3.1037}

\bibitem[{Goudfrooij {et~al.}(2017)Goudfrooij, Girardi, \&
  Correnti}]{goudfrooij_extended_2017}
Goudfrooij, P., Girardi, L., \& Correnti, M. 2017, The Astrophysical Journal,
  846, 22, \dodoi{10.3847/1538-4357/aa7fb7}

\bibitem[{Goudfrooij {et~al.}(2011)Goudfrooij, Puzia, Chandar, \&
  Kozhurina-Platais}]{goudfrooij_population_2011}
Goudfrooij, P., Puzia, T.~H., Chandar, R., \& Kozhurina-Platais, V. 2011, The
  Astrophysical Journal, 737, 4, \dodoi{10.1088/0004-637X/737/1/4}

\bibitem[{Goudfrooij {et~al.}(2009)Goudfrooij, Puzia, Kozhurina-Platais, \&
  Chandar}]{goudfrooij_population_2009}
Goudfrooij, P., Puzia, T.~H., Kozhurina-Platais, V., \& Chandar, R. 2009, The
  Astronomical Journal, 137, 4988, \dodoi{10.1088/0004-6256/137/6/4988}

\bibitem[{Harris {et~al.}(2020)Harris, Millman, van~der Walt, Gommers,
  Virtanen, Cournapeau, Wieser, Taylor, Berg, Smith, Kern, Picus, Hoyer, van
  Kerkwijk, Brett, Haldane, del R{\'\i}o, Wiebe, Peterson, G{\'e}rard-Marchant,
  Sheppard, Reddy, Weckesser, Abbasi, Gohlke, \& Oliphant}]{numpy}
Harris, C.~R., Millman, K.~J., van~der Walt, S.~J., {et~al.} 2020, Nature, 585,
  357, \dodoi{10.1038/s41586-020-2649-2}

\bibitem[{He {et~al.}(2022)He, Sun, Li, Shao, Zhong, Chen, Grijs, Tang, Qin, \&
  Randriamanakoto}]{he_role_2022}
He, C., Sun, W., Li, C., {et~al.} 2022, The Astrophysical Journal, 938, 42,
  \dodoi{10.3847/1538-4357/ac8b08}

\bibitem[{Heggie(1975)}]{10.1093/mnras/173.3.729}
Heggie, D.~C. 1975, Monthly Notices of the Royal Astronomical Society, 173,
  729, \dodoi{10.1093/mnras/173.3.729}

\bibitem[{Hills(1975)}]{1975AJ.....80..809H}
Hills, J.~G. 1975, \aj, 80, 809, \dodoi{10.1086/111815}

\bibitem[{Hills \& Day(1976)}]{hills_stellar_1976}
Hills, J.~G., \& Day, C.~A. 1976, Astrophysical Letters, 17, 87.
\newblock \url{https://ui.adsabs.harvard.edu/abs/1976ApL....17...87H}

\bibitem[{Hunter(2007)}]{4160265}
Hunter, J.~D. 2007, Computing in Science \& Engineering, 9, 90,
  \dodoi{10.1109/MCSE.2007.55}

\bibitem[{Hurley {et~al.}(2000)Hurley, Pols, \&
  Tout}]{10.1046/j.1365-8711.2000.03426.x}
Hurley, J.~R., Pols, O.~R., \& Tout, C.~A. 2000, Monthly Notices of the Royal
  Astronomical Society, 315, 543, \dodoi{10.1046/j.1365-8711.2000.03426.x}

\bibitem[{Hurley {et~al.}(2002)Hurley, Tout, \&
  Pols}]{10.1046/j.1365-8711.2002.05038.x}
Hurley, J.~R., Tout, C.~A., \& Pols, O.~R. 2002, Monthly Notices of the Royal
  Astronomical Society, 329, 897, \dodoi{10.1046/j.1365-8711.2002.05038.x}

\bibitem[{Iwasawa {et~al.}(2020)Iwasawa, Namekata, Nitadori, Nomura, Wang,
  Tsubouchi, \& Makino}]{10.1093/pasj/psz133}
Iwasawa, M., Namekata, D., Nitadori, K., {et~al.} 2020, Publications of the
  Astronomical Society of Japan, 72, \dodoi{10.1093/pasj/psz133}

\bibitem[{Iwasawa {et~al.}(2016)Iwasawa, Tanikawa, Hosono, Nitadori, Muranushi,
  \& Makino}]{10.1093/pasj/psw053}
Iwasawa, M., Tanikawa, A., Hosono, N., {et~al.} 2016, Publications of the
  Astronomical Society of Japan, 68, \dodoi{10.1093/pasj/psw053}

\bibitem[{Kamann {et~al.}(2021)Kamann, Bastian, Usher, Cabrera-Ziri, \&
  Saracino}]{kamann_exploring_2021}
Kamann, S., Bastian, N., Usher, C., Cabrera-Ziri, I., \& Saracino, S. 2021,
  Monthly Notices of the Royal Astronomical Society, 508, 2302,
  \dodoi{10.1093/mnras/stab2643}

\bibitem[{Kamann {et~al.}(2019)Kamann, Bastian, Gossage, Baade, Cabrera-Ziri,
  Da Costa, de Mink, Georgy, Giesers, Göttgens, Hilker, Husser, Lardo,
  Larsen, Mackey, Martocchia, Mucciarelli, Platais, Roth, Salaris, Usher, \&
  Yong}]{10.1093/mnras/stz3583}
Kamann, S., Bastian, N., Gossage, S., {et~al.} 2019, Monthly Notices of the
  Royal Astronomical Society, 492, 2177, \dodoi{10.1093/mnras/stz3583}

\bibitem[{Korntreff {et~al.}(2012)Korntreff, Kaczmarek, \&
  Pfalzner}]{korntreff_towards_2012}
Korntreff, C., Kaczmarek, T., \& Pfalzner, S. 2012, Astronomy and Astrophysics,
  543, A126, \dodoi{10.1051/0004-6361/201118019}

\bibitem[{Kroupa(1995{\natexlab{a}})}]{10.1093/mnras/277.4.1507}
Kroupa, P. 1995{\natexlab{a}}, Monthly Notices of the Royal Astronomical
  Society, 277, 1507, \dodoi{10.1093/mnras/277.4.1507}

\bibitem[{Kroupa(1995{\natexlab{b}})}]{10.1093/mnras/277.4.1491}
---. 1995{\natexlab{b}}, Monthly Notices of the Royal Astronomical Society,
  277, 1491, \dodoi{10.1093/mnras/277.4.1491}

\bibitem[{Kroupa(2001)}]{10.1046/j.1365-8711.2001.04022.x}
---. 2001, Monthly Notices of the Royal Astronomical Society, 322, 231,
  \dodoi{10.1046/j.1365-8711.2001.04022.x}

\bibitem[{K\"upper {et~al.}(2011)K\"upper, Maschberger, Kroupa, \&
  Baumgardt}]{10.1111/j.1365-2966.2011.19412.x}
K\"upper, A. H.~W., Maschberger, T., Kroupa, P., \& Baumgardt, H. 2011, Monthly
  Notices of the Royal Astronomical Society, 417, 2300,
  \dodoi{10.1111/j.1365-2966.2011.19412.x}

\bibitem[{Leigh {et~al.}(2015)Leigh, Giersz, Marks, Webb, Hypki, Heinke,
  Kroupa, \& Sills}]{Leigh2015TheSO}
Leigh, N. W.~C., Giersz, M., Marks, M., {et~al.} 2015, Monthly Notices of the
  Royal Astronomical Society, 446, 226

\bibitem[{Li {et~al.}(2016)Li, de~Grijs, Bastian, Deng, Niederhofer, \&
  Zhang}]{li_tight_2016}
Li, C., de~Grijs, R., Bastian, N., {et~al.} 2016, Monthly Notices of the Royal
  Astronomical Society, 461, 3212, \dodoi{10.1093/mnras/stw1491}

\bibitem[{Li {et~al.}(2014{\natexlab{a}})Li, de~Grijs, \&
  Deng}]{li_exclusion_2014}
Li, C., de~Grijs, R., \& Deng, L. 2014{\natexlab{a}}, Nature, 516, 367,
  \dodoi{10.1038/nature13969}

\bibitem[{Li {et~al.}(2017)Li, de~Grijs, Deng, \& Milone}]{Li_2017}
Li, C., de~Grijs, R., Deng, L., \& Milone, A.~P. 2017, The Astrophysical
  Journal, 834, 156, \dodoi{10.3847/1538-4357/834/2/156}

\bibitem[{Li {et~al.}(2014{\natexlab{b}})Li, Grijs, \&
  Deng}]{li_not-so-simple_2014}
Li, C., Grijs, R.~d., \& Deng, L. 2014{\natexlab{b}}, The Astrophysical
  Journal, 784, 157, \dodoi{10.1088/0004-637X/784/2/157}

\bibitem[{Li {et~al.}(2019)Li, Sun, Grijs, Deng, Wang, Cordoni, \&
  Milone}]{li_extended_2019}
Li, C., Sun, W., Grijs, R.~d., {et~al.} 2019, The Astrophysical Journal, 876,
  65, \dodoi{10.3847/1538-4357/ab15d2}

\bibitem[{Li {et~al.}(2023)Li, Liu, Zhang, Tian, Fu, Li, \&
  Yan}]{li_stellar_2023}
Li, J., Liu, C., Zhang, Z.-Y., {et~al.} 2023, Nature, 613, 460,
  \dodoi{10.1038/s41586-022-05488-1}

\bibitem[{Mackey {et~al.}(2008)Mackey, Nielsen, Ferguson, \&
  Richardson}]{Mackey_2008}
Mackey, A.~D., Nielsen, P.~B., Ferguson, A. M.~N., \& Richardson, J.~C. 2008,
  The Astrophysical Journal, 681, L17, \dodoi{10.1086/590343}

\bibitem[{Maeder \& Meynet(2000)}]{doi:10.1146/annurev.astro.38.1.143}
Maeder, A., \& Meynet, G. 2000, Annual Review of Astronomy and Astrophysics,
  38, 143, \dodoi{10.1146/annurev.astro.38.1.143}

\bibitem[{Marino {et~al.}(2018{\natexlab{a}})Marino, Milone, Casagrande,
  Przybilla, Balaguer-Núñez, Criscienzo, Serenelli, \&
  Vilardell}]{Marino_2018}
Marino, A.~F., Milone, A.~P., Casagrande, L., {et~al.} 2018{\natexlab{a}}, The
  Astrophysical Journal Letters, 863, L33, \dodoi{10.3847/2041-8213/aad868}

\bibitem[{Marino {et~al.}(2018{\natexlab{b}})Marino, Przybilla, Milone, Costa,
  D’Antona, Dotter, \& Dupree}]{marino_different_2018}
Marino, A.~F., Przybilla, N., Milone, A.~P., {et~al.} 2018{\natexlab{b}}, The
  Astronomical Journal, 156, 116, \dodoi{10.3847/1538-3881/aad3cd}

\bibitem[{McLaughlin \& Marel(2005)}]{mclaughlin_resolved_2005}
McLaughlin, D.~E., \& Marel, R. P. v.~d. 2005, The Astrophysical Journal
  Supplement Series, 161, 304, \dodoi{10.1086/497429}

\bibitem[{McLaughlin \& van~der Marel(2005)}]{McLaughlin_2005}
McLaughlin, D.~E., \& van~der Marel, R.~P. 2005, The Astrophysical Journal
  Supplement Series, 161, 304, \dodoi{10.1086/497429}

\bibitem[{Milone {et~al.}(2009)Milone, Bedin, Piotto, \&
  Anderson}]{milone_multiple_2009}
Milone, A.~P., Bedin, L.~R., Piotto, G., \& Anderson, J. 2009, Astronomy \&
  Astrophysics, 497, 755, \dodoi{10.1051/0004-6361/200810870}

\bibitem[{Milone {et~al.}(2016)Milone, Marino, D'Antona, Bedin, Da~Costa,
  Jerjen, \& Mackey}]{milone_multiple_2016}
Milone, A.~P., Marino, A.~F., D'Antona, F., {et~al.} 2016, Monthly Notices of
  the Royal Astronomical Society, 458, 4368, \dodoi{10.1093/mnras/stw608}

\bibitem[{Milone {et~al.}(2015)Milone, Bedin, Piotto, Marino, Cassisi, Bellini,
  Jerjen, Pietrinferni, Aparicio, \& Rich}]{10.1093/mnras/stv829}
Milone, A.~P., Bedin, L.~R., Piotto, G., {et~al.} 2015, Monthly Notices of the
  Royal Astronomical Society, 450, 3750, \dodoi{10.1093/mnras/stv829}

\bibitem[{Milone {et~al.}(2017)Milone, Marino, D'Antona, Bedin, Piotto, Jerjen,
  Anderson, Dotter, Criscienzo, \& Lagioia}]{milone_multiple_2017}
Milone, A.~P., Marino, A.~F., D'Antona, F., {et~al.} 2017, Monthly Notices of
  the Royal Astronomical Society, 465, 4363, \dodoi{10.1093/mnras/stw2965}

\bibitem[{Milone {et~al.}(2018)Milone, Marino, Di~Criscienzo, D'Antona, Bedin,
  Da~Costa, Piotto, Tailo, Dotter, Angeloni, Anderson, Jerjen, Li, Dupree,
  Granata, Lagioia, Mackey, Nardiello, \& Vesperini}]{10.1093/mnras/sty661}
Milone, A.~P., Marino, A.~F., Di~Criscienzo, M., {et~al.} 2018, Monthly Notices
  of the Royal Astronomical Society, 477, 2640, \dodoi{10.1093/mnras/sty661}

\bibitem[{Milone {et~al.}(2022)Milone, Cordoni, Marino, D'Antona, Bellini,
  Di~Criscienzo, Dondoglio, Lagioia, Langer, Legnardi, Libralato, Baumgardt,
  Bettinelli, Cavecchi, de~Grijs, Deng, Hastings, Li, Mohandasan, Renzini,
  Vesperini, Wang, Ziliotto, Carlos, Costa, Dell'Agli, Di~Stefano, Jang,
  Martorano, Simioni, Tailo, \& Ventura}]{milone_hubble_2022}
Milone, A.~P., Cordoni, G., Marino, A.~F., {et~al.} 2022, Hubble {Space}
  {Telescope} survey of {Magellanic} {Cloud} star clusters. {Photometry} and
  astrometry of 113 clusters and early results,  arXiv.
\newblock \url{http://arxiv.org/abs/2212.07978}

\bibitem[{Moeckel \& Bonnell(2009)}]{10.1111/j.1365-2966.2009.15499.x}
Moeckel, N., \& Bonnell, I.~A. 2009, Monthly Notices of the Royal Astronomical
  Society, 400, 657, \dodoi{10.1111/j.1365-2966.2009.15499.x}

\bibitem[{Namekata {et~al.}(2018)Namekata, Iwasawa, Nitadori, Tanikawa,
  Muranushi, Wang, Hosono, Nomura, \& Makino}]{10.1093/pasj/psy062}
Namekata, D., Iwasawa, M., Nitadori, K., {et~al.} 2018, Publications of the
  Astronomical Society of Japan, 70, \dodoi{10.1093/pasj/psy062}

\bibitem[{Niederhofer {et~al.}(2015)Niederhofer, Georgy, Bastian, \&
  Ekström}]{niederhofer_apparent_2015}
Niederhofer, F., Georgy, C., Bastian, N., \& Ekström, S. 2015, Monthly Notices
  of the Royal Astronomical Society, 453, 2070, \dodoi{10.1093/mnras/stv1791}

\bibitem[{Oh {et~al.}(2015)Oh, Kroupa, \& Pflamm-Altenburg}]{Oh_2015}
Oh, S., Kroupa, P., \& Pflamm-Altenburg, J. 2015, The Astrophysical Journal,
  805, 92, \dodoi{10.1088/0004-637X/805/2/92}

\bibitem[{Parker {et~al.}(2013)Parker, Wright, Goodwin, \&
  Meyer}]{10.1093/mnras/stt2231}
Parker, R.~J., Wright, N.~J., Goodwin, S.~P., \& Meyer, M.~R. 2013, Monthly
  Notices of the Royal Astronomical Society, 438, 620,
  \dodoi{10.1093/mnras/stt2231}

\bibitem[{Plummer(1911)}]{10.1093/mnras/71.5.460}
Plummer, H.~C. 1911, Monthly Notices of the Royal Astronomical Society, 71,
  460, \dodoi{10.1093/mnras/71.5.460}

\bibitem[{Rivinius {et~al.}(2013)Rivinius, Carciofi, \&
  Martayan}]{rivinius_classical_2013}
Rivinius, T., Carciofi, A.~C., \& Martayan, C. 2013, The Astronomy and
  Astrophysics Review, 21, 69, \dodoi{10.1007/s00159-013-0069-0}

\bibitem[{R\"oser \& Schilbach(2019)}]{R2019}
R\"oser, S., \& Schilbach, E. 2019, A\&A, 627, A4,
  \dodoi{10.1051/0004-6361/201935502}

\bibitem[{{R\"oser, S.} {et~al.}(2011){R\"oser, S.}, {Schilbach, E.},
  {Piskunov, A. E.}, {Kharchenko, N. V.}, \& {Scholz, R.-D.}}]{R2011}
{R\"oser, S.}, {Schilbach, E.}, {Piskunov, A. E.}, {Kharchenko, N. V.}, \&
  {Scholz, R.-D.} 2011, A\&A, 531, A92, \dodoi{10.1051/0004-6361/201116948}

\bibitem[{Sana {et~al.}(2012)Sana, de~Mink, de~Koter, Langer, Evans, Gieles,
  Gosset, Izzard, Bouquin, \& Schneider}]{sana}
Sana, H., de~Mink, S.~E., de~Koter, A., {et~al.} 2012, Science, 337, 444,
  \dodoi{10.1126/science.1223344}

\bibitem[{{Spitzer}(1987)}]{1987degc.book.....S}
{Spitzer}, L. 1987, {Dynamical evolution of globular clusters}

\bibitem[{Sun {et~al.}(2019)Sun, Li, Deng, \&
  de~Grijs}]{sun_tidal-locking-induced_2019}
Sun, W., Li, C., Deng, L., \& de~Grijs, R. 2019, The Astrophysical Journal,
  883, 182, \dodoi{10.3847/1538-4357/ab3cd0}

\bibitem[{Tout {et~al.}(1997)Tout, Aarseth, Pols, \&
  Eggleton}]{10.1093/mnras/291.4.732}
Tout, C.~A., Aarseth, S.~J., Pols, O.~R., \& Eggleton, P.~P. 1997, Monthly
  Notices of the Royal Astronomical Society, 291, 732,
  \dodoi{10.1093/mnras/291.4.732}

\bibitem[{Virtanen {et~al.}(2020)Virtanen, Gommers, Oliphant, Haberland, Reddy,
  Cournapeau, Burovski, Peterson, Weckesser, Bright, van~der Walt, Brett,
  Wilson, Millman, Mayorov, Nelson, Jones, Kern, Larson, Carey, Polat, Feng,
  Moore, VanderPlas, Laxalde, Perktold, Cimrman, Henriksen, Quintero, Harris,
  Archibald, Ribeiro, Pedregosa, van Mulbregt, Vijaykumar, Bardelli, Rothberg,
  Hilboll, Kloeckner, Scopatz, Lee, Rokem, Woods, Fulton, Masson,
  H{\"a}ggstr{\"o}m, Fitzgerald, Nicholson, Hagen, Pasechnik, Olivetti, Martin,
  Wieser, Silva, Lenders, Wilhelm, Young, Price, Ingold, Allen, Lee, Audren,
  Probst, Dietrich, Silterra, Webber, Slavi{\v c}, Nothman, Buchner, Kulick,
  Sch{\"o}nberger, de~Miranda~Cardoso, Reimer, Harrington, Rodr{\'\i}guez,
  Nunez-Iglesias, Kuczynski, Tritz, Thoma, Newville, K{\"u}~mmerer,
  Bolingbroke, Tartre, Pak, Smith, Nowaczyk, Shebanov, Pavlyk, Brodtkorb, Lee,
  McGibbon, Feldbauer, Lewis, Tygier, Sievert, Vigna, Peterson, More, Pudlik,
  Oshima, Pingel, Robitaille, Spura, Jones, Cera, Leslie, Zito, Krauss,
  Upadhyay, Halchenko, V{\'a}zquez-Baeza, \& Contributors}]{Scipy}
Virtanen, P., Gommers, R., Oliphant, T.~E., {et~al.} 2020, Nature Methods, 17,
  261, \dodoi{10.1038/s41592-019-0686-2}

\bibitem[{von Zeipel(1924)}]{10.1093/mnras/84.9.665}
von Zeipel. 1924, Monthly Notices of the Royal Astronomical Society, 84, 665,
  \dodoi{10.1093/mnras/84.9.665}

\bibitem[{Wang {et~al.}(2022{\natexlab{a}})Wang, Langer, Schootemeijer, Milone,
  Hastings, Xu, Bodensteiner, Sana, Castro, Lennon, {et~al.}}]{wang2022stellar}
Wang, C., Langer, N., Schootemeijer, A., {et~al.} 2022{\natexlab{a}}, Nature
  Astronomy, 6, 480

\bibitem[{Wang(2020)}]{wang_survival_2020}
Wang, L. 2020, Monthly Notices of the Royal Astronomical Society, 491, 2413,
  \dodoi{10.1093/mnras/stz3179}

\bibitem[{Wang {et~al.}(2020)Wang, Iwasawa, Nitadori, \&
  Makino}]{10.1093/mnras/staa1915}
Wang, L., Iwasawa, M., Nitadori, K., \& Makino, J. 2020, Monthly Notices of the
  Royal Astronomical Society, 497, 536, \dodoi{10.1093/mnras/staa1915}

\bibitem[{Wang {et~al.}(2018)Wang, Kroupa, \&
  Jerabkova}]{10.1093/mnras/sty2232}
Wang, L., Kroupa, P., \& Jerabkova, T. 2018, Monthly Notices of the Royal
  Astronomical Society, 484, 1843, \dodoi{10.1093/mnras/sty2232}

\bibitem[{Wang {et~al.}(2022{\natexlab{b}})Wang, Tanikawa, \&
  Fujii}]{wang_impact_2022}
Wang, L., Tanikawa, A., \& Fujii, M.~S. 2022{\natexlab{b}}, Monthly Notices of
  the Royal Astronomical Society, 509, 4713, \dodoi{10.1093/mnras/stab3255}

\bibitem[{Yang {et~al.}(2013)Yang, Bi, Meng, \& Liu}]{yang_effects_2013}
Yang, W., Bi, S., Meng, X., \& Liu, Z. 2013, The Astrophysical Journal, 776,
  112, \dodoi{10.1088/0004-637X/776/2/112}

\bibitem[{Yang {et~al.}(2021)Yang, Li, Grijs, \& Deng}]{yang_spatial_2021}
Yang, Y., Li, C., Grijs, R.~d., \& Deng, L. 2021, The Astrophysical Journal,
  912, 27, \dodoi{10.3847/1538-4357/abec4b}

\bibitem[{{Zahn}(1975)}]{1975A&A....41..329Z}
{Zahn}, J.~P. 1975, \aap, 41, 329

\end{thebibliography}
\bibliographystyle{aasjournal}

\end{CJK*}
\end{document}